\documentclass[journal,twoside,web]{ieeecolor}
\usepackage{tuffc}
\usepackage{cite}
\usepackage{amsmath,amssymb,amsfonts}
\usepackage{algorithmic}
\usepackage{graphicx}
\usepackage{textcomp}
\usepackage{subfigure}
\usepackage{xcolor}
\def\BibTeX{{\rm B\kern-.05em{\sc i\kern-.025em b}\kern-.08em
    T\kern-.1667em\lower.7ex\hbox{E}\kern-.125emX}}
\graphicspath{{pics/}{figs/}}
\begin{document}
\title{3D Ultrafast Shear Wave Absolute Vibro-Elastography using a Matrix Array Transducer}
\author{Hoda S. Hashemi, \IEEEmembership{Member, IEEE}, Shahed K. Mohammed, \IEEEmembership{Member, IEEE}, Qi Zeng, \IEEEmembership{Member, IEEE}, Reza Zahiri Azar, \IEEEmembership{Member, IEEE}, Robert N. Rohling, \IEEEmembership{Fellow, IEEE} and Septimiu E. Salcudean, \IEEEmembership{Fellow, IEEE}
\thanks{\textcolor{black}{This work was supported by the CA Laszlo Chair in Biomedical
Engineering held by Professor Salcudean, and the Natural Sciences and Engineering Research
Council of Canada (NSERC).}}
\thanks{Hoda Hashemi, Shahed Mohammed, and Qi Zeng are with the Electrical and Computer Engineering Department, University of British Columbia, Vancouver, BC, Canada (e-mails: hoda,shahedkm,qizeng@ece.ubc.ca). }
\thanks{Reza Zahiri is with the DarkVision Technologies Inc., North Vancouver, BC, Canada (e-mails: reza.zahiri@darkvisiontech.com).}
\thanks{Robert Rohling is with the Electrical and Computer Engineering, and Mechanical Department, University of British Columbia, Vancouver, BC, Canada (e-mails: rohling@ece.ubc.ca).}
\thanks{Septimiu Salcudean is with the Electrical and Computer Engineering, and Biomedical Engineering Department, University of British Columbia, Vancouver, BC, Canada (e-mails: tims@ece.ubc.ca).}}

\maketitle

\begin{abstract}
Real-time ultrasound imaging plays an important role in ultrasound-guided interventions. 3D imaging provides more spatial information compared to conventional 2D frames by considering the volumes of data. One of the main bottlenecks of 3D imaging is the long data acquisition time which reduces practicality and can introduce artifacts from unwanted patient or sonographer motion. 

This paper introduces the first \textcolor{black}{shear wave absolute vibro-elastography (S-WAVE) method with \textit{real-time volumetric acquisition}} using a matrix array transducer. In S-WAVE, an external vibration source generates mechanical vibrations inside the tissue. The tissue motion is then estimated and used in solving a wave equation inverse problem to provide the tissue elasticity.
A matrix array transducer is used with a Verasonics ultrasound machine and frame rate of 2000~volumes/s to acquire 100 radio frequency (RF) volumes in 0.05~s. 
\textcolor{black}{Using plane wave (PW) and compounded diverging wave (CDW) imaging methods, we estimate} axial, lateral and elevational displacements over 3D volumes. The curl of the displacements is used with local frequency estimation to estimate elasticity in the acquired volumes.\\
Ultrafast acquisition extends substantially the possible S-WAVE excitation frequency range, now up to 800~Hz, enabling new tissue modeling and characterization. 
The method was validated on three \textcolor{black}{homogeneous} liver fibrosis phantoms \textcolor{black}{and on \textcolor{black}{four} different inclusions within a heterogeneous phantom}. 
\textcolor{black}{The homogeneous phantom results show less than 8\% (PW) and 5\% (CDW) difference between the manufacturer values and the corresponding estimated values} over a frequency range of 80~Hz to~800 Hz.
\textcolor{black}{The estimated elasticity values for the heterogeneous phantom at 400~Hz excitation frequency show average errors of 9\% (PW) and 6\% (CDW) compared to the provided average values by MRE. Furthermore, both imaging methods were able to detect the inclusions within the elasticity volumes.}\\
An {\em ex vivo} study on a bovine liver sample shows \textcolor{black}{less than 11\% (PW) and 9\% (CDW) difference between the estimated elasticity ranges by the proposed method and the elasticity ranges provided by MRE and ARFI.}
\end{abstract}


\section{Introduction}
\label{sec:introduction}
\IEEEPARstart{R}{eal-time} ultrasound imaging \textcolor{black}{is not only useful for diagnostic B-mode imaging, but also is} used in monitoring, targeting, controlling, and assessing treatment response during medical procedures such as thermotherapy~\cite{liu2009real,lewis2015thermometry}, ultrasound-guided biopsy~\cite{brock2012impact,moe2020transrectal}, MRI-US fusion targeted biopsy\cite{miyagawa2010real,valerio2015detection,pinto2011magnetic}, and molecular imaging of cancer~\cite{hyun2020nondestructive,perera2020real}.
Minimally invasive thermal therapies such as RF ablation and high-intensity-focused ultrasound (HIFU) are treatment options for malignancies that elevate tissue temperature leading to tumor destruction~\cite{dodd2000minimally}. \textcolor{black}{In these medical procedures}, noninvasive temperature estimation is necessary to protect against over-heating and coagulation necrosis in surrounding tissue. Furthermore, temperature estimation should be performed in real-time due to fast switching of the heating source~\cite{wi2015real}. However, conventional frame-rate ultrasound is not suitable for real-time temperature estimation for the following reasons: First, it suffers from inherent tissue deformations due to breathing and cardiac \textcolor{black}{motion}. \textcolor{black}{Tissue deformations} lead to echo and spectral shifts which cause masking of the temperature-induced shifts. Second, \textcolor{black}{temporal} undersampling of the tissue displacement fields at typical frame rates of ultrasound imaging can produce artifacts. Therefore, high-frame-rate RF acquisition is needed to capture the transients of tissue motion in response to pulsed HIFU or on/off control of RF ablation to estimate the temperature change~\cite{liu2009real}. 

\textcolor{black}{In a similar direction,} elastography \textcolor{black}{is} a quantitative method \textcolor{black}{that} can display the stiffness of both abdominal tumours and thermal lesions and therefore can be used for noninvasive guiding and assessing thermal therapies~\cite{bharat2005monitoring,van2010intra,hashemi2017global}. Furthermore, real-time elastography-guided biopsy improves cancer lesion detection compared to conventional gray scale ultrasound guidance~\cite{brock2012impact,van2014comparison,postema2015multiparametric}. \textcolor{black}{For example,} Shear Wave Absolute Vibro-Elastography (S-WAVE) is an elastography method~\cite{abeysekera2015vibro,abeysekera2017swave} where a multi-frequency external excitation source generates sinusoidal vibrations inside the medium. 
By measuring the tissue motion between ultrasound images and solving the inverse problem of the wave equation~\cite{mohammed2022model}, the absolute elasticity of the tissue can be estimated over a tissue volume~\cite{zeng2020three,zeng2022multifrequency}. Volumetric image acquisition in S-WAVE takes of the order of seconds to acquire. For example, a large volume liver scan can take 12 s to acquire~\cite{zeng2020three}, requiring patients to hold their breath for at least that long, which is not always possible for elderly or paediatric patients~\cite{riccabona2003potential}. 
As an alternative solution to make the imaging faster, previous work~\cite{hashemi2022ultrafast} combined ultrafast elastography with the S-WAVE system to acquire ultrasound images in quasi-real-time. 2D ultrasound frames were acquired \textcolor{black}{using a motorized (wobbler) transducer} at different plane locations to form volumes of data. Although 100 ultrasound volumes were collected in 1.5-2~s, unwanted tissue motion caused by breathing \textcolor{black}{and unintentional transducer motion by the operator can still affect the results by introducing image artifacts. One of the drawbacks of the mechanical wobbler approach is the need for synchronization between plane imaging}, transducer motion and the excitation frequency period.

One application of real-time imaging is in the assessment of liver fibrosis, caused by progressing nonalcoholic fatty liver disease (NAFLD), which affects more than 25\% of the global population~\cite{loomba2013global}. Furthermore, the prevalence of NAFLD among children \textcolor{black}{under the age of 18} is between 3 to 10\%, rising up to 40 to 70\% among obese children~\cite{bellentani2010epidemiology}. Magnetic resonance elastography (MRE) is used as a non-invasive imaging solution without the need for invasive biopsies. However, MRE is relatively inaccessible. It has been shown that tissue elasticity is significantly correlated with fibrosis stage~\cite{foucher2006diagnosis}. So, ultrasound elastography can be used as a \textcolor{black}{more accessible} screening and diagnosis tool where the tissue elasticity is considered as a biomarker for staging the fibrosis. S-WAVE has been successfully used to estimate liver stiffness in both healthy subjects and patients~\cite{zeng2020three}. Therefore, \textcolor{black}{a fast} 3D S-WAVE can enhance the liver stiffness exams by substantially improving the examination time, eliminating the effect of unwanted motions due to breathing, and removing any artifacts from high effective rate sector-based imaging by using plane waves and insonifying the entire imaging region at the same time.

\textcolor{black}{This paper proposes a 3D imaging method with \textit{real-time 3D image acquisition} using a matrix array transducer. We call this method 3D Ultrafast S-WAVE. We use frame rates of 2000 volumes/s to acquire 100 RF volumes with a total data collection time of 0.05~s. The elasticity of the tissue is estimated by using a 3D, curl-based, wave equation model. Furthermore, we substantially extend the range of excitation frequencies relative to previous S-WAVE studies~\cite{shao2021breast,zeng2020three,hashemi2022ultrafast}, thereby enabling a more comprehensive tissue modeling and characterization over a wider frequency range~\cite{nightingale2015derivation,aichele2021fluids}. }

The proposed method is validated on three homogeneous liver fibrosis phantoms with different elasticity values representing the range of healthy livers to those with fibrosis and early cirrhosis. An \textit{ex vivo} study using a bovine liver sample is performed to show the results of the method on real tissue. The same sample was tested by MRE and an acoustic radiation force impulse (ARFI) method~\cite{deng2016ultrasonic} to compare the range of elasticity values provided by different techniques. The \textcolor{black}{advances} of this work over the previous S-WAVE \textcolor{black}{research} are as follows:
\begin{enumerate}
\item Introducing the first 3D Ultrafast S-WAVE method with \textit{real-time} image acquisition.
\item Extending the excitation frequency range of S-WAVE (up to 800 Hz) enabling more comprehensive tissue modeling,
\item Enabling new applications for the S-WAVE method by using a small footprint matrix array transducer.  
\end{enumerate}

The rest of this paper is organized as follows: Section~\ref{sec:methods} explains the imaging setup, displacements, phasors, curl, and elasticity estimation. Section~\ref{sec:results} validates the method using CIRS phnatoms and shows the results of the \textit{ex vivo} data, and compares the range of elasticity values with MRE and an ARFI technique~\cite{deng2016ultrasonic} where plane waves are also utilized. The purpose of the \textit{ex vivo} study is to investigate the consistency of the elasticity ranges measured with different methods. Possible future work and study limitations are discussed in Section~\ref{sec:discussion}. Finally, our conclusions are presented in Section~\ref{sec:conclusion}.

\begin{figure}
\centering
\includegraphics[width=0.49\columnwidth]{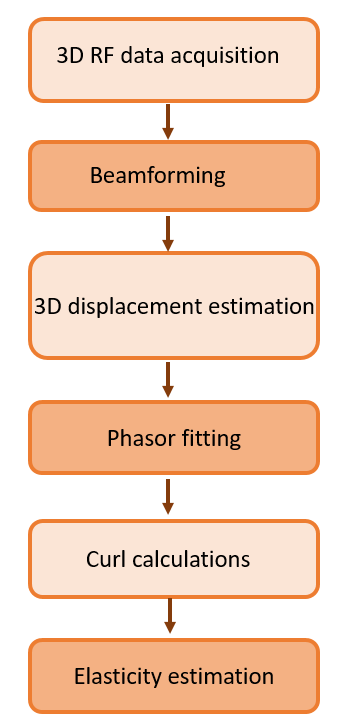}\\
\caption{\textcolor{black}{Schematic diagram of the 3D Ultrafast S-WAVE method.}}
\label{fig:method}
\end{figure}

\begin{figure}
\centering
\includegraphics[width=0.7\columnwidth]{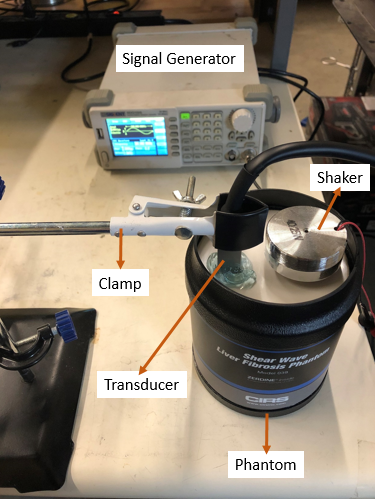}\\
\caption{\textcolor{black}{Ultrafast S-WAVE setup. A signal generator is connected to the external shaker that generates a sinusoidal excitation inside the phantom. The matrix array transducer is fixed on top of the phantom using a transducer clamp and is also connected to a Verasonics ultrasound machine.} }
\label{fig:phantomSetup}
\end{figure}

\section{Methods}
\label{sec:methods}

\subsection{Imaging Method}
\label{subsec:ImagingMethod}
\textcolor{black}{A schematic diagram of the method is shown in Figure~\ref{fig:method}. The overall hardware and each step are explained in details as follows:}

\subsubsection{Overall Hardware}
\label{subsec:OverallHardware}
\textcolor{black}{The experiment\textcolor{black}{al} setup is shown in Figure~\ref{fig:phantomSetup}. A signal generator is connected to a voice-coil shaker (50mm 20W Vibration Resonance Speaker, FTVOGUE, PR China), placed on top of the medium to generate sinusoidal vibrations inside the tissue. RF echo data is acquired at a sampling rate of 20 MHz using a Vantage 256 ultrasound system (Verasonics Inc., Kirkland, WA) with a 5 MHz center frequency matrix array (Imasonic, Voray sur l’Ognon, France).} \textcolor{black}{The matrix array transducer has 256 elements placed on a $16\times 16$ quadratic footprint with no dead elements. The size of the transducer footprint is $1 \times 1$~cm$^2$ with pitch values of 0.6 mm for both lateral and elevational directions.} 

\begin{figure*}
\centering
\subfigure[3D view]{\includegraphics[width=0.7\columnwidth]{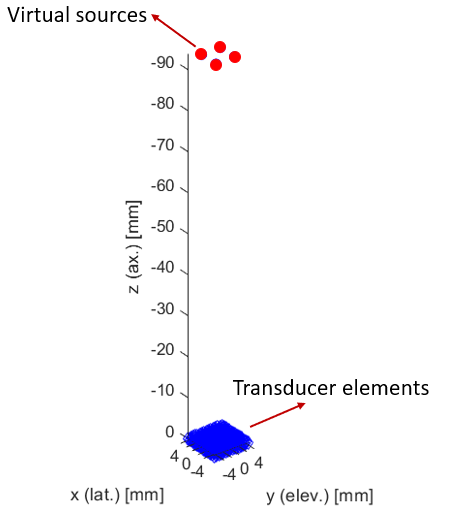}}
\subfigure[Axial-lateral plane]{\includegraphics[width=0.7\columnwidth]{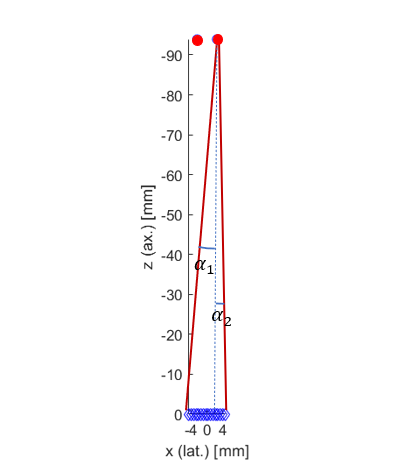}}\\
\subfigure[Lateral-elevational plane]{\includegraphics[width=0.7\columnwidth]{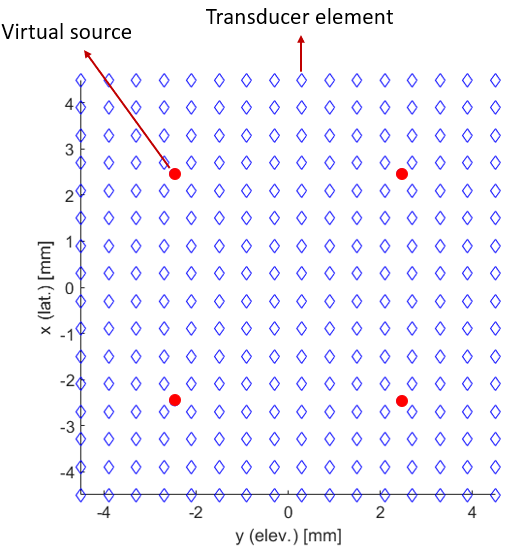}}
\subfigure[Delays~(s)]{\includegraphics[width=0.95\columnwidth]{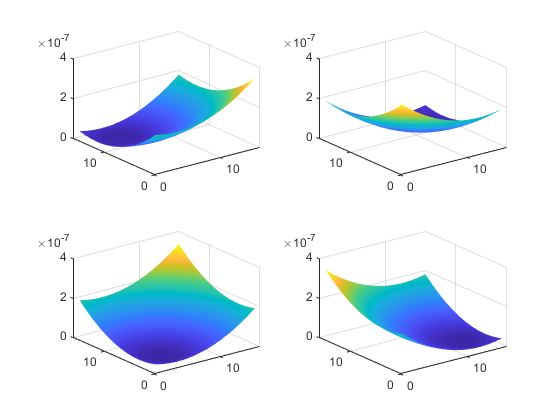}}
\caption{\textcolor{black}{The location of virtual sources in the compounded diverging waves sequence. (a)-(c) show four virtual sources with spacing of 0.5~cm are located at a distance equal to 9.4~cm behind the surface of the transducer and used to obtain an approximately $\theta=6^{\circ}$ angle aperture in the lateral and elevational directions ($\alpha_1\approx4.5^{\circ}$, $\alpha_2\approx1.5^{\circ}$). The calculated delays for each virtual source are shown in (d).}}
\label{fig:virtualSources}
\end{figure*}

\subsubsection{Imaging Sequences \& beamforming}
\label{subsec:ImagingSequences}
\textcolor{black}{Two different imaging sequences are used in this work: first, a plane wave (PW) sequence, and second, a compounded diverging wave (CDW) sequence. For PW, single plane wave transmit and receive events with $0^\circ$ insonifications (or equivalently, delays of 0~ms) are used to acquire 100 RF volumes with the volume rate of 2000 volumes/s. For CDW, the diverging waves are transmitted from four virtual sources located behind the surface of the transducer for compounding emissions (Figure~\ref{fig:virtualSources} (a)).} 

\textcolor{black}{The number of virtual sources depends on the application and can be optimized based on several factors such as volume rate and imaging depth. Provost et al.~\cite{provost20143d} used one virtual source behind the transducer to track the propagation of the shear wave to maintain a high frame rate. Papadacci et al.~\cite{papadacci2014high} showed a large improvement in focusing quality of the point spread function (PSF) by increasing the number of virtual sources from one to three. In this work, we place four virtual sources on a symmetric $2 \times 2$ grid to form a virtual array behind the transducer. The pulse repetition frequency (PRF) is set to 8000 Hz resulting in an overall volume rate of 2000 volumes/s after compounding. This PRF allows for a maximum imaging depth of 9.6~cm which is calculated as $1540/(8000\times2$). To increase the imaging depth, the PRF or the number of virtual sources should be decreased. The recommended maximum steering angle is determined  by the probe pitch~\cite{wilcox2018quantification}, and is $4.5^{\circ}$.
In order to achieve the maximum steering angle, the virtual sources are placed at a distance of 9.4~cm behind the transducer} \textcolor{black}{surface which provides an approximately $6^{\circ}$ field of view in the lateral and elevational directions as shown in Figure~\ref{fig:virtualSources} (b) ($\alpha_1\approx4.5^{\circ}$, $\alpha_2\approx1.5^{\circ}$).
For each source, a set of delays is calculated for all aperture elements as the full aperture of the transducer was used to generate \textcolor{black}{the} diverging beams (Figure~\ref{fig:virtualSources} (d)).
The emissions from \textcolor{black}{the virtual} sources are carried out sequentially, and the RF data is recorded for each element of the entire transducer. 
After collecting RF data, the volumes are beamformed by applying the delay-and-sum algorithm for each virtual source, and then coherently compounding \textcolor{black}{beamformed volumes of four transmit events} to produce a high-quality volume. 
For beamforming of both sequences, a 3D grid in Cartesian coordinates with resolution $(\lambda/8, P/2, P/2)$ in the axial, lateral, and elevational directions, respectively, is defined where $\lambda$ is the ultrasound wavelength and $P$ is the pitch of the transducer in the lateral and elevation directions. In beamforming, lower F-number values provide better lateral resolution whereas higher values result in an image with fewer artifacts. In this work, the F-number is adjusted to 1 where an acceptable lateral resolution and image quality are visually observed in the B-mode images. 
The fan-shaped field of view for the compounded diverging wave sequence can be represented by defining a Polar coordinate beamforming grid. However, to maintain consistency with the plane wave sequence, an identical beamforming grid in the Cartesian coordinate system is used for both sequences.} \textcolor{black}{A GPU-based} \textcolor{black}{implementation of the delay-and-sum algorithm is then used to beamform the RF volumes~\cite{perrot2021so}} \textcolor{black}{which takes 0.03 s to beamform and compound four RF volumes of size $7\times1\times1$~cm$^3$ on a 3.6-GHz Intel Core i7 computer.}

\textcolor{black}{There are several trade-offs between volume rate, number of virtual sources, virtual pitch (spacing between the virtual sources), and contrast and resolution.  
A previous study showed as the number of sources is increased, the contrast and resolution improve\cite{provost20143d}, but the volume rate decreases. High volume rate is required for applications such as shear wave imaging or ultrafast Doppler imaging. 
A large virtual pitch would increase the lateral resolution, but would also increase the grating lobes. As shown in Figure~\ref{fig:virtualSources} (c), in this work, the spacing between the virtual sources is considered as half of the size of the transducer aperture (0.5~cm)~\cite{papadacci2014high,grondin2017cardiac}.
For the compounded diverging wave sequence, similar to the plane wave imaging sequence, a total of 100 compounded RF volumes are collected to be used in displacement estimation.}

\subsubsection{Displacement Phasor Estimation}
\label{subsec:DispEstimation}
\textcolor{black}{The three-dimensional relative shear-wave displacement field is estimated using the GLUE3D algorithm~\cite{hashemi20203d} where a regularized cost function incorporates the data intensity and displacement continuity to be optimized in 3D. The displacement between two consequent volumes are tiny due to a high frame rate which can make the final estimation noisy. The algorithm uses the displacement obtained by a 3D normalized cross correlation (NCC) method~\cite{azar2010sub} as an initial guess and starts to fine-tune it by optimizing a quadratic cost function~\cite{hashemi20203d}. It was shown in a previous study that the algorithm is more robust to noise and provides smoother displacements compared to the conventional NCC method~\cite{hashemi2017global}.} \textcolor{black}{Similar to other displacement} \textcolor{black}{estimation methods, the GLUE3D algorithm has adjustable parameters ($\alpha_{1,2}, \beta_{1,2}, \gamma_{1,2}, t$) that can be set at the beginning of the ultrasound exam based on the target organ. The parameters $\alpha_{1,2}, \beta_{1,2}, \gamma_{1,2}$ regulate the spatial displacement continuity in the axial, lateral, and elevational directions, respectively, and the parameter $t$ regulates the temporal displacement continuity throughout the data volumes. Higher parameter values lead to smoother displacement maps which can be inspected visually, and are supported by lower STD values. In all experiments,} \textcolor{black}{the coefficients of the GLUE3D algorithm are adjusted to the recommended values explained in~\cite{hashemi20203d} such that $\alpha_{1,2}=\beta_{1,2}=\gamma_1=200$, $\gamma_2 = 0.5$, and $t = 0.001$.} \textcolor{black}{Changing the GLUE3D coefficients by 100\% changes the elasticity values by less than 2\% for both phantom and \textit{ex vivo} data. So, the results are not sensitive to these parameter values.} \textcolor{black}{The displacement estimation is followed by fitting displacement phasors using a least-squares solution~\cite{abeysekera2016three}} \textcolor{black}{to provide axial, lateral, and elevational displacement phasor volumes.}

\subsubsection{Curl Calculation}
\label{subsec:CurlCalculation}

\textcolor{black}{Axial, lateral, and elevational displacement phasor \textcolor{black}{volumes} are used in the 3D curl calculations.} \textcolor{black}{A fine 3D grid of points is defined in all three directions with the step size of 0.05~mm which is approximately equal to the axial resolution of the \textcolor{black}{beamforming grid}. All the three phasor volumes are projected to this fine grid of points where the resolution is the same in all directions.} \textcolor{black}{The curl of a 3D volume can be estimated as follows:\\
$\nabla\times \textbf{D} = (\frac{\partial D_{z}}{\partial y}-\frac{\partial D_{y}}{\partial z}) \hat{x} + (\frac{\partial D_{x}}{\partial z}-\frac{\partial D_{z}}{\partial x}) \hat{y} + (\frac{\partial D_{y}}{\partial x}-\frac{\partial D_{x}}{\partial y}) \hat{z}$,
where $D$ is the 3D displacement phasor, x, y and z are axial, lateral, and elevational directions, respectively. The curl is calculated for each point on the grid resulting in a 3D vector for that point. Therefore, by performing the curl operator for all the points within the volume, three volumes corresponding to axial, lateral and elevational directions will be achieved. The partial derivatives are calculated}  \textcolor{black}{with an step size of 0.05~mm} \textcolor{black}{using the central difference method within the volume, and forward difference method on the boundaries. Therefore, the calculations are more accurate within the volume.} \textcolor{black}{In order to alleviate the noise of the curl operation, a 3D Gaussian filter with $\sigma = 2$ has been applied to the curl volumes and the samples in the curl volumes are averaged in a window of size $4\times4\times4$ samples to make a coarser grid of points with the resolution of 0.2 mm to decrease the volume size and save the computational time in reconstruction.}  
\textcolor{black}{The advantage of reconstruction using the 3D curl of displacements over reconstruction with no curl (using axial, lateral, and elevational displacement phasors) was shown in previous work\cite{hashemi2022ultrafast}} \textcolor{black}{and also is examined \textcolor{black}{briefly} in this study.}

\subsubsection{Elasticity Estimation}
\label{subsec:ElasticityEstimation}

\textcolor{black}{The three volumes of curl displacements phasors are used in the} \textcolor{black}{local frequency estimation (LFE) algorithm to estimate the local spatial frequency of the shear wave propagation} \textcolor{black}{pattern at the excitation frequency by combining local estimates of instantaneous frequency over several scales~\cite{knutsson1994local,manduca1996image}. Each curl displacement phasor volume is processed independently by LFE. First, each volume is translated into the spatial frequency domain (k-space) using the Fast Fourier Transform. The phasors are filtered using a sixth-order Butterworth bandpass filter corresponding to the range of 0.5 kPa to 40 kPa to alleviate any potential noise from the curl differentiation process. A directional filter bank consisting of six k-space directional filters along the axial, lateral and elevational directions ($\pm x$, $\pm y$ and $\pm z$) are applied to the 3D curl of phasor volumes. The directional filters can reduce the elasticity reconstruction artifacts due to the shear wave reflection from the boundaries~\cite{deffieux2011effects}.} \textcolor{black}{Then, each directional component is passed through a radial filter bank consisting of log-normal filters corresponding to 11 central frequencies spaced an octave apart, from $\frac{1}{2^{10}}$ to $\frac{1}{2^0}$.} 
\textcolor{black}{The magnitude ratio between two consecutive log-normal filters provides a narrow-band estimate of local frequency or multiple angular wave-numbers. A weighted summation of these estimates provides the final angular wave-number $k$, where the magnitudes of the log-normal filters are used as the weights. The angular wave-number $k$ is transformed to the elasticity $E$ as follows:
\begin{equation}
\label{eq:shearWaveSpeed2E}
E = 3 \rho {C_s}^2 = 3 \rho ({f_e \lambda})^2 = 3 \rho \left(\frac{2\pi f_e}{k}\right)^2
\end{equation}
where $\rho$ is the density of the tissue and is assumed to be 1000 $\frac{Kg}{m^3}$, $C_s$ is the local shear wave speed, $f_e$ is the excitation frequency, and $\lambda$ is the shear wave spatial wavelength.}

\begin{figure*}
\begin{center}
\subfigure[Ultrafast S-WAVE]{\includegraphics[width=0.31\textwidth] {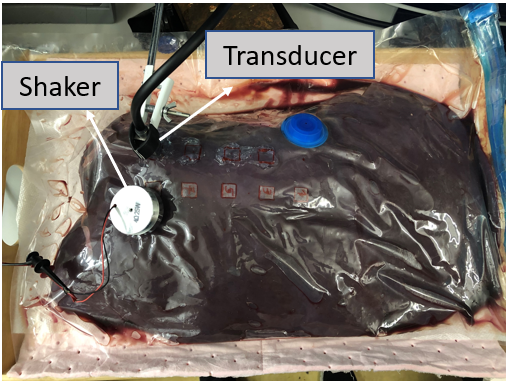}}
\subfigure[MRE]{\includegraphics[width=0.31\textwidth] {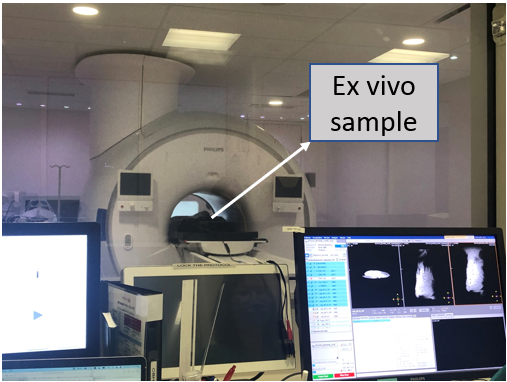}}
\subfigure[ARFI]{\includegraphics[width=0.31\textwidth] {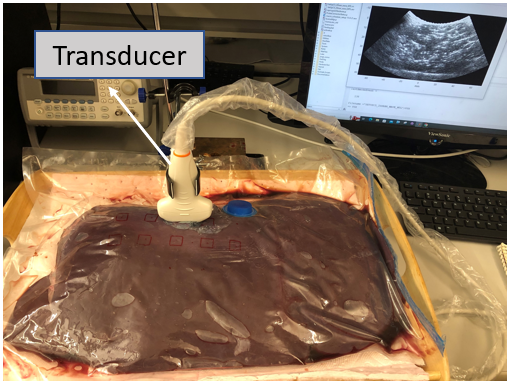}}
\caption{\textit{Ex vivo} experiment setup for the bovine liver sample. (a) shows \textcolor{black}{the Ultrafast S-WAVE setup}. The experimental setup for \textcolor{black}{the MRE} and ARFI are shown in (b) and (c), respectively. Seven locations are marked on the sample to be used for the Ultrafast S-WAVE and ARFI measurements. The marked locations are also known and visible in MR images by attaching vitamin E capsules.}
\label{fig:exvivoSetup}
\end{center}
\end{figure*}

\subsection{3D Ultrafast S-WAVE Experiments}
\label{subsec:SWAVEExperiments}
\textcolor{black}{The proposed method was applied to three CIRS homogeneous liver fibrosis phantoms, and a heterogeneous CIRS phantom with four different inclusions and an \textit{ex vivo} bovine liver sample as explained as follows:} 
 
\subsubsection{Homogeneous Phantom}
\label{subsec:HomogeneousPhantom}
\textcolor{black}{Volumetric RF data \textcolor{black}{were} collected using both of \textcolor{black}{PW and CDW} imaging sequences} \textcolor{black}{from three homogeneous phantoms with the elasticity values of 1.8 kPa, 7.8 kPa, and 17.5 kPa with a precision of $\pm 4\%$, representing a healthy liver, early fibrosis, and progressive fibrosis/early cirrhosis respectively~\cite{foucher2006diagnosis}. The elasticity values change with respect to the excitation frequency and the imaging depth is 7~cm. An excitation frequency range of [80 - 800] Hz is used such that the higher frequencies are considered for the stiffer phantom to be consistent with the provided elasticity values by the manufacturer. The highest frequency used in this study (800 Hz) meets the Nyquist criterion (lower than $f_{Nyquist} = 1000~Hz$) considering the volume rate of 2000 volumes/s.}
\textcolor{black}{The shear wave spatial wavelength can be calculated from Eq. \ref{eq:shearWaveSpeed2E} as: 
\begin{equation}
\label{eq:spatialWaveLength}
\lambda = \frac{1}{f_e} \sqrt{\frac{E}{3}}
\end{equation}
where $f_e$ is the frequency of the excitation in [Hz] and $E$ is the elasticity of the tissue in [kPa] assuming the tissue has a density of 1000 $[\frac{kg}{m^3}]$.  
The excitation frequencies were selected such that the spatial wavelength is less than 1~cm to have at least one full spatial wavelength of the shear waves in all three directions.
This selection will be discussed further in the \ref{sec:discussion} Section. The estimated elasticity values are validated using CIRS phantom manufacturer elasticity values.}

\subsubsection{Heterogeneous Phantom}
\label{subsec:HeterogeneousPhantom}
\textcolor{black}{RF data is acquired from four different regions within a heterogeneous phantom (Model 049, CIRS Inc., Norfolk, VA, USA). Each imaging region contains a spherical inclusion. The diameter of the inclusions is 1~cm. \textcolor{black}{Both of proposed imaging sequences are} used to image the heterogeneous phantom. The elasticity range of the inclusions reported by the manufacturer are as follows: 5.5 to 7 kPa for the inclusion 1, 10 to 12.5 kPa for the inclusion 2, 32 to 41 kPa for the inclusion 3, 56 to 71 \textcolor{black}{kPa} for the inclusion 4, and 17.5 to 22 kPa for the background. Two inclusions are softer than the background and the other two are stiffer than the background. The excitation \textcolor{black}{frequency of 400~Hz is used to vibrate the inclusions which} allows to have at least one spatial wavelength in each direction (Equation~\ref{eq:spatialWaveLength}) and also meets the Nyquist criterion.} 
\textcolor{black}{Before the main data collection, the locations of the four spherical inclusions inside the heterogeneous phantom are confirmed and marked on the phantom with B-mode images collected with a L12-3v linear array transducer (Verasonics Inc., Kirkland, WA). For the main data collection, the matrix array transducer is then placed on top of the phantom where each inclusion is located.}

\subsubsection{\textit{Ex Vivo} Bovine Liver}
\label{subsubsec:method_exvivo}
\textcolor{black}{The \textit{ex vivo} data was collected from bovine liver tissue. The experimental setup for the Ultrafast S-WAVE is shown in Figure~\ref{fig:exvivoSetup} (a). The sample dimensions are $48 \times 28$~cm with a depth of $6$~cm in the axial direction, and a weight of 6~kg. All data acquisitions are
performed at room temperature ($20^\circ$ C). The sample was placed in a vacuum bag. Seven locations with a distance of 4~cm are marked on the sample. The matrix array is placed on each marked location on the sample to provide a volume of size $1\times1\times6$~cm$^3$ (depth is 6~cm) of elasticity for each location. The measurements are performed for each location by averaging within a $5\times5\times10$~mm$^3$ ROI to avoid boundary artifacts and at the depth of 3~cm to 4~cm which is approximately at the center of the sample depth (6~cm). Two excitation frequencies of 60 Hz and 100 Hz are used to vibrate the tissue which are within} \textcolor{black}{the range of excitation frequencies previously used for the bovine liver ex vivo studies~\cite{ormachea2016shear}. 
For the \textit{ex vivo} experiment, as there is no ground truth for the actual elasticity} \textcolor{black}{value of the sample, we also estimate the elasticity of the sample using MRE and ARFI methods to investigate the range of the elasticity values estimated by different methods.}

\subsection{MRE Validation}
\label{subsec:MREValidation}
\textcolor{black}{The MRE exam setup for the \textit{ex vivo} experiment is shown in Figure~\ref{fig:exvivoSetup}(b). MRE is performed using two excitation frequencies of 60 Hz and 100 Hz which are within the range of previously used excitation frequencies for the bovine liver {\em ex vivo} studies~\cite{ormachea2016shear} \textcolor{black}{and also used in the {\em ex vivo} measurements by the Ultrafast S-WAVE method.} MRE scans the whole sample and provides an elasticity volume for all the regions within the sample. In order to perform the measurements at the same locations used previously for the Ultrafast S-WAVE experiment, seven vitamin E capsules are attached on seven marked locations which are visible in MR images.}
\textcolor{black}{To compare the results with the proposed method, the obtained elasticity volumes from the Ultrafast S-WAVE are then registered on the MR images by having the vitamin E capsules locations in MR images. This step is performed manually through the Slicer Image Registration Toolkit~\cite{kikinis20143d} with an accuracy of approximately 0.5 cm. The final elasticity volume is calculated for the whole liver sample region, and the measurements are performed at each marked location at the depth of 3 to 4 cm within a $5\times5\times10$~mm$^3$ ROI to be consistent with the ROI size in the Ultrafast S-WAVE {\em ex vivo} experiment.}

\subsection{ARFI Validation}
\label{subsec:ARFIValidation}
\textcolor{black}{The experimental setup is shown in Figure~\ref{fig:exvivoSetup} (c) where a C5-2v curved linear array is connected to a Verasonics ultrasound machine and an implementation of the ARFI method as explained in \cite{deng2016ultrasonic} is used to collect and process the data. The ARFI method reports the elasticity at a single point on the middle line of the transducer and the measurement depth should be specified before the start of the ARFI exam. For the ARFI data collection, we placed the transducer on top of each marked locations explained earlier to provides a value for each location at a focal depth of 3.5~cm.} \textcolor{black}{The phase velocity estimation is performed by averaging over the axial depth of field (DOF) window which corresponds to an approximately 7.7 mm window at a frequency of 3.5 MHz, and defined by $8F^2\lambda$ around the focal depth of the push beam where F is the F-number and $\lambda$ is the wavelength of the push beam~\cite{deng2016ultrasonic}. The displacement tracking and phase velocity estimation are performed at the same focal depth of 3.5~cm. So, the averaging region would be approximately between the depth of 3~cm to 4~cm which is consistent with the other \textit{ex vivo} experiments. The average phase velocities for 60 Hz and 100 Hz are converted to the elasticity values using Equation~\ref{eq:shearWaveSpeed2E}.}

\begin{figure}
\centering
\subfigure[Axial displacement phasor]{\includegraphics[width=1\columnwidth]{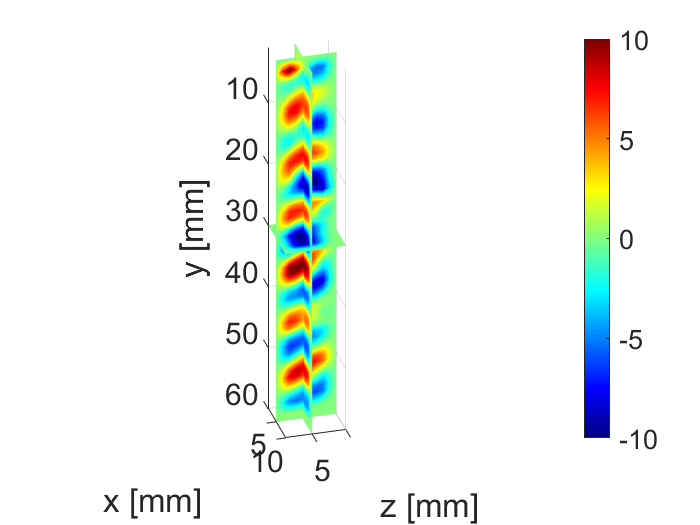}}
\subfigure[Lateral displacement phasor]{\includegraphics[width=1\columnwidth]{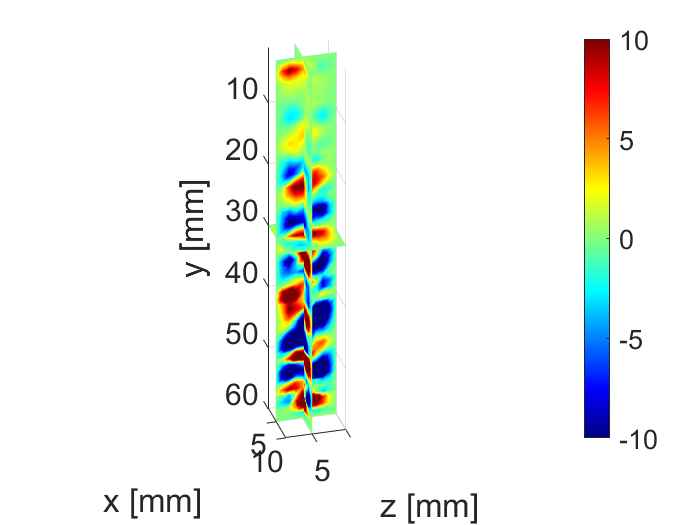}}
\subfigure[Elevational displacement phasor]{\includegraphics[width=1\columnwidth]{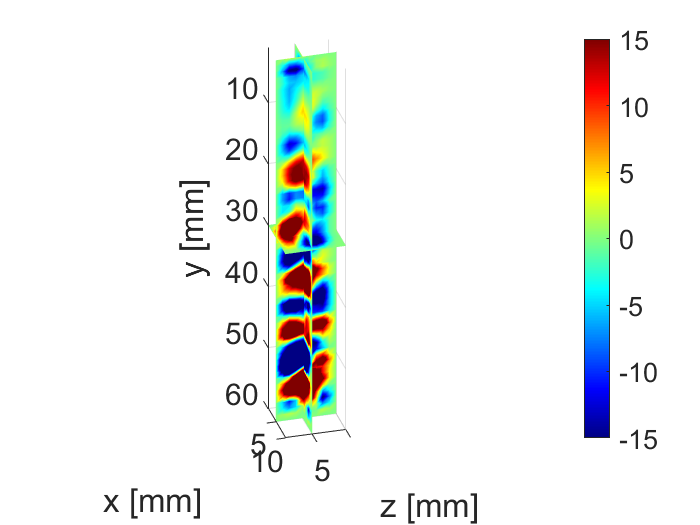}}
\caption{\textcolor{black}{Displacement phasors for the 17.5 kPa phantom. (a)-(c) show the axial, lateral, and elevational phasor volumes for the excitation frequency of 300 Hz. The value bar is in [$\mu$m].}}
\label{fig:phantomPhasors}
\end{figure}

\begin{figure*}
\begin{center}
\subfigure[phantom 1 (1.8~kPa)]{\includegraphics[width=1.4\columnwidth]{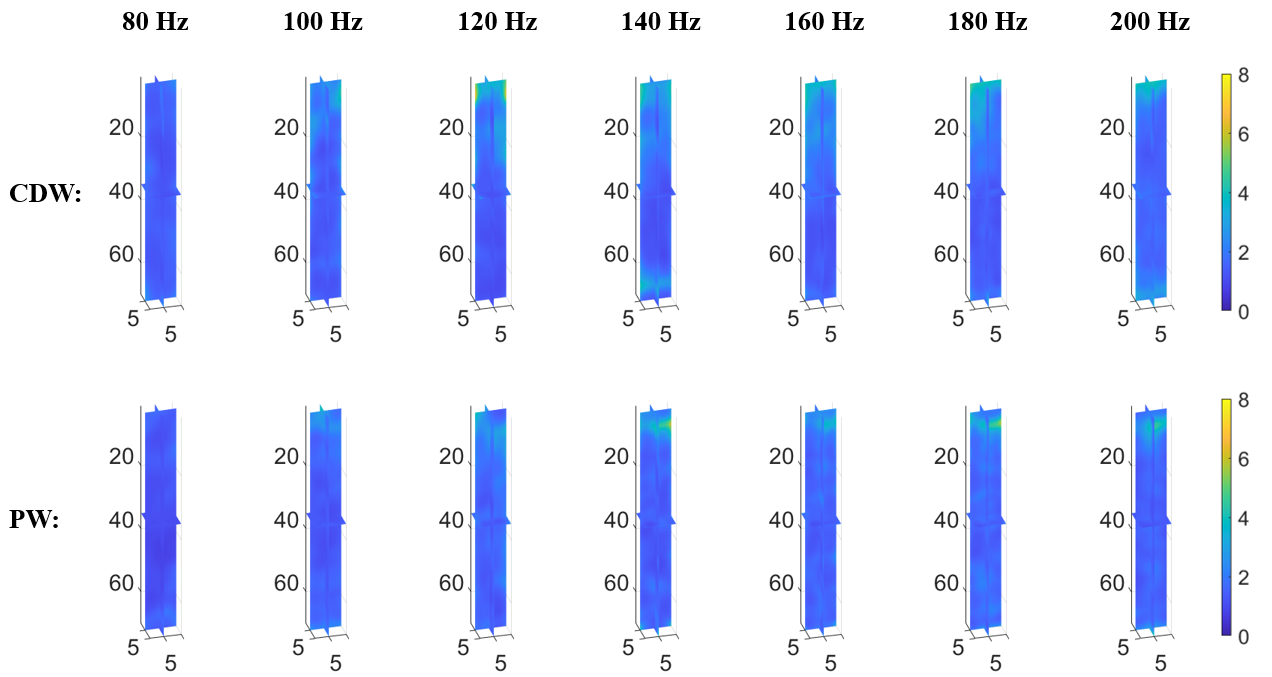}}\\
\subfigure[phantom 2 (7.8~kPa)]{\includegraphics[width=1.4\columnwidth]{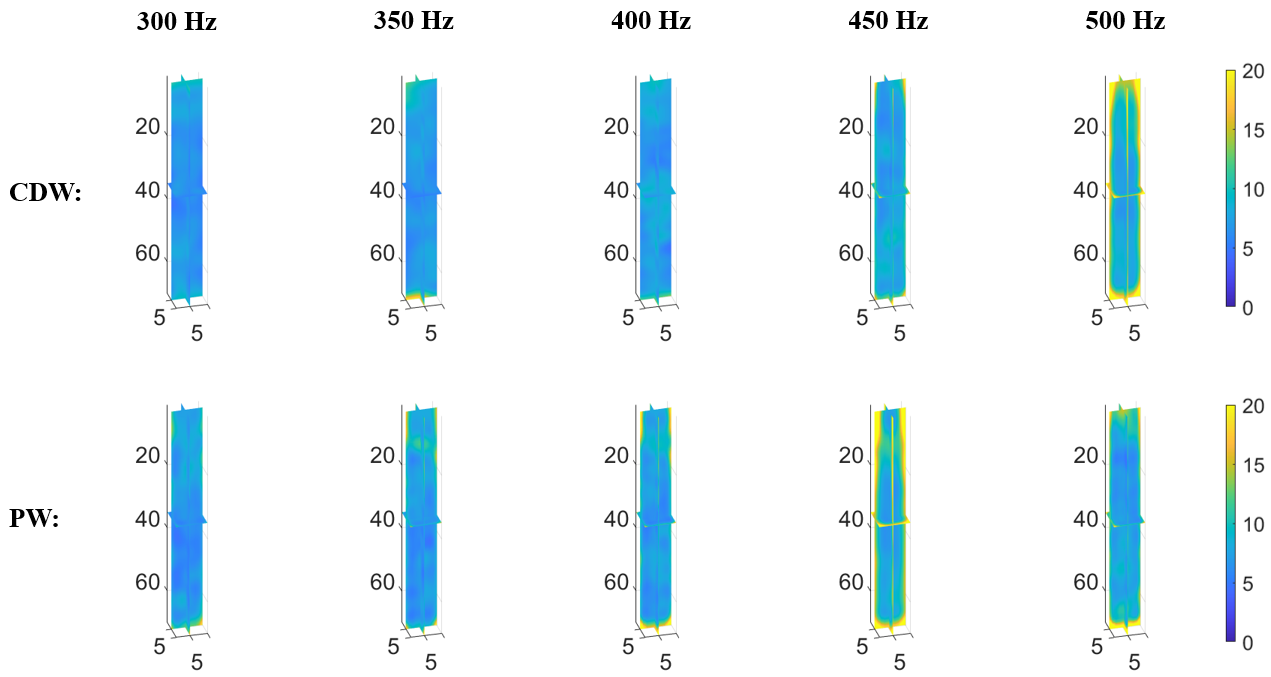}}\\
\subfigure[phantom 3 (17.5~kPa)]{\includegraphics[width=1.4\columnwidth]{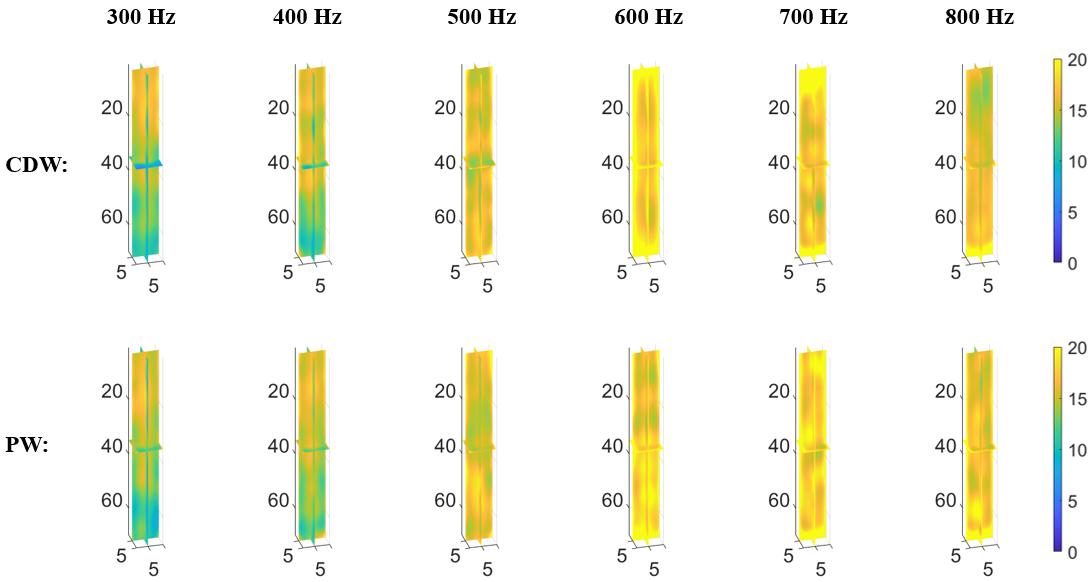}}\\
\end{center}
\caption{\textcolor{black}{Elasticity volumes for the homogeneous liver fibrosis CIRS phantoms. (a)-(c) show the elasticity volumes for each excitation frequency for 1.8~kPa, 7.8~kPa, and 17.5~kPa phantoms, respectively. \textcolor{black}{For each excitation frequency, the results of the plane wave (PW) sequence and also compounded diverging wave (CDW) sequence are shown.} The values of the midline of the volumes between 10 and 60 mm depth are used for the elasticity calculations. The value bar is in [kPa].}}
\label{fig:phantomEvolumes}
\end{figure*}

\section{Results}
\label{sec:results}
\textcolor{black}{In this section the results of the phantom and \textit{ex vivo} experiments are presented.} 

\subsection{Homogeneous Phantom}

\begin{figure}
\begin{center}
\subfigure[\textcolor{black}{1.8 kPa Phantom}]{\includegraphics[width=0.45\textwidth] {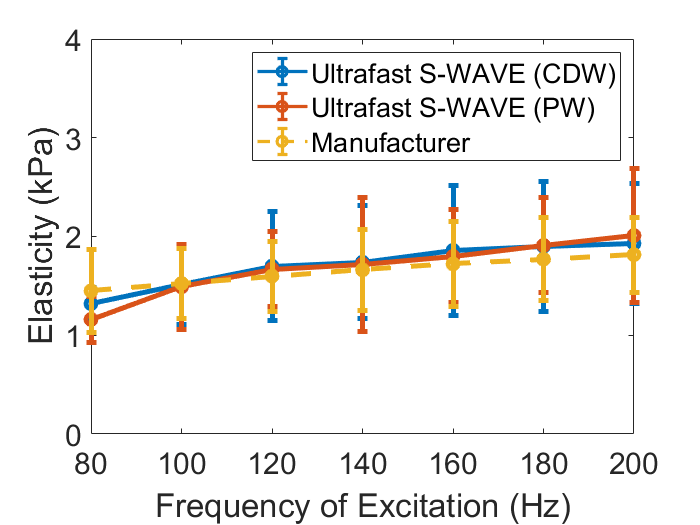}}
\subfigure[7.8 kPa Phantom ]{\includegraphics[width=0.45\textwidth] {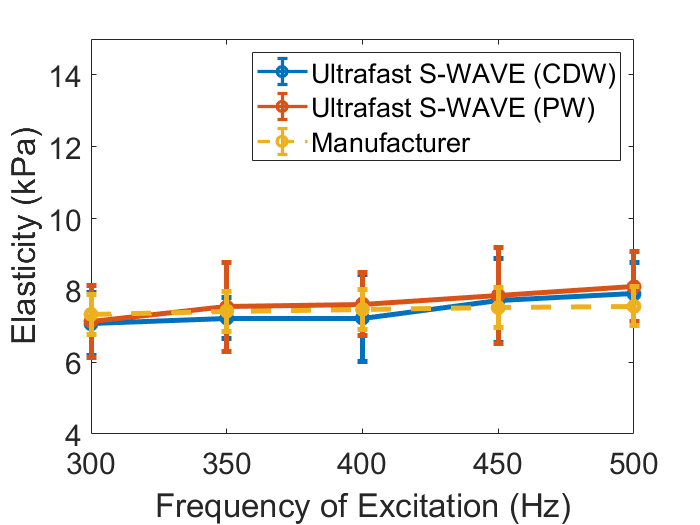}}
\subfigure[17.4 kPa Phantom ]{\includegraphics[width=0.45\textwidth] {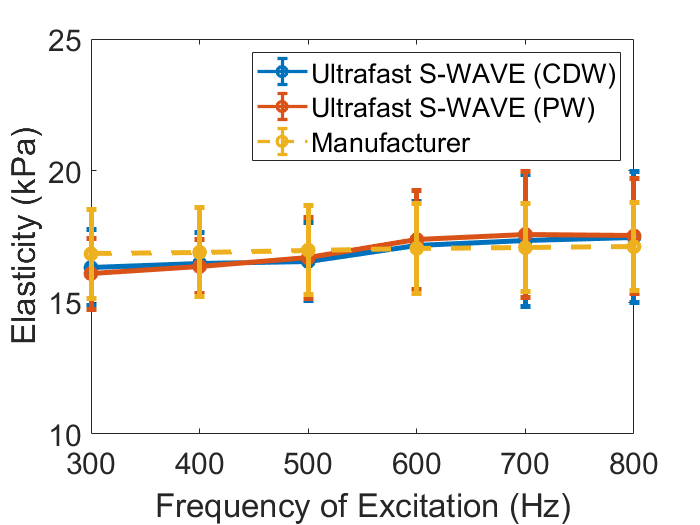}}
\caption{Estimated elasticity values (average $\pm$ STD) for three homogeneous CIRS phantoms. \textcolor{black}{The average elasticity and STD values for each frequency are calculated along the midline of the transducer, starting at 1 cm from the face of the transducer to 1 cm from the bottom of the image. The yellow bars represent the reported range by the manufacturer. The red and blue dots show the average values by the PW and CDW sequences, respectively. The average error between the average elasticity values reported by the manufacturer and the Ultrafast S-WAVE method are less than 8\% and 5\% for PW and CDW, respectively, for all excitation frequencies.}}
\label{fig:phantomE}
\end{center}
\end{figure}

\textcolor{black}{Displacement phasor volumes for the 17.5 kPa phantom using the compounded diverging wave sequence are shown in Figure~\ref{fig:phantomPhasors}. 
(a)-(c) show axial, lateral, and elevational phasors at the excitation frequency of 300 Hz. The value bar is shown in $\mu$m.} \textcolor{black}{Figure~\ref{fig:phantomEvolumes} shows the elasticity volumes of all homogeneous phantoms for each excitation frequency for the PW and CDW imaging sequences.} 
\textcolor{black}{Figure~\ref{fig:phantomE} shows the estimated average elasticity values and standard deviations (STD) at each excitation frequency as well as the corresponding manufacturer values for each phantom.} \textcolor{black}{The manufacturer elasticity values are averaged along a window around an ARFI focal point in the axial direction only, which provides an average value for a single point~\cite{deng2016ultrasonic}. To be consistent with the manufacturer measurements and also avoid boundary artifacts associated with the LFE algorithm~\cite{manduca2018waveguide}, for all elasticity volumes, the average elasticity and STD} \textcolor{black}{values are calculated along the midline of the transducer, starting 1 cm from the face of the transducer to 1 cm from the bottom of the image. However,}
\textcolor{black}{the corresponding value for the manufacturer is given for a single point (averaged along a window in axial direction~\cite{deng2016ultrasonic}), and therefore, results in lower STD at some frequencies.}
\textcolor{black}{To be consistent with the manufacturer excitation frequency range, higher excitation frequencies are used for the stiffer phantom. The estimated elasticity values for all phantoms fall within the manufacturer’s elasticity range for all excitation frequencies.} \textcolor{black}{The average difference between the manufacturer values for each frequency and the estimated ones by the proposed method is less than 8$\%$ for the PW sequence and less than 6$\%$ for the CDW sequence.}

\begin{figure}
\begin{center}
\subfigure[3D reconstruction with axial displacement volume]{\includegraphics[width=0.45\textwidth] {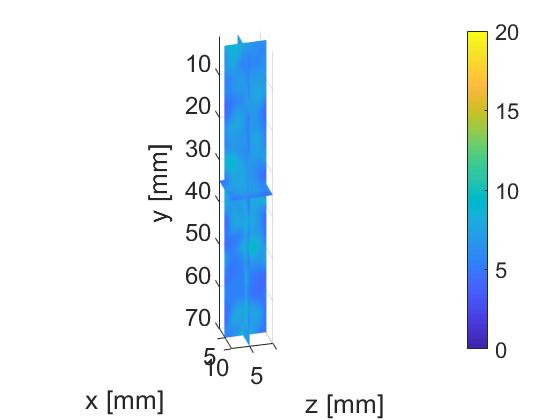}}
\subfigure[3D reconstruction without curl]{\includegraphics[width=0.45\textwidth] {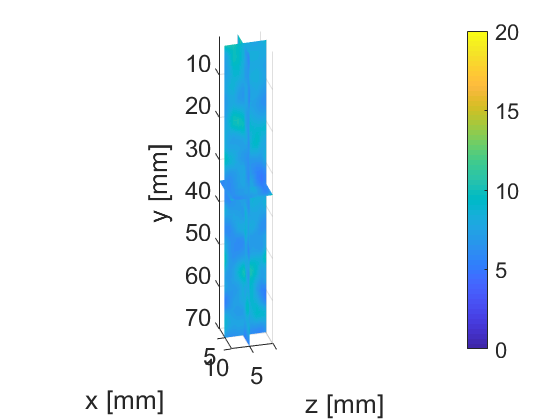}}
\subfigure[3D reconstruction with curl]{\includegraphics[width=0.45\textwidth] {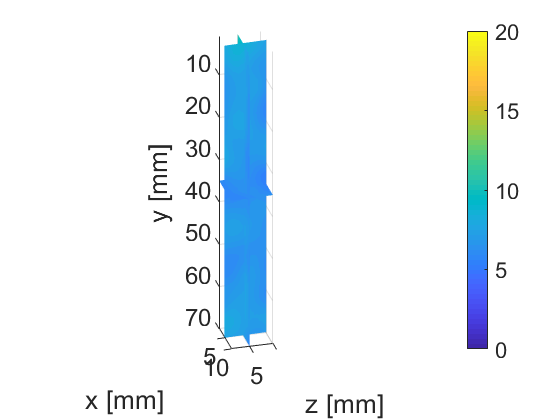}}
\caption{\textcolor{black}{Elasticity volumes for the 7.8 kPa homogeneous phantom with an excitation frequency of 350 Hz. (a) shows the elasticity volume using 3D reconstruction with only the axial displacement phasor volume (1D over 3D). (b) shows the 3D reconstruction result using axial, lateral, and elevational phasors without using the curl (3D over 3D). (c) shows the result of the 3D reconstruction using the curl of axial, lateral and elevational displacement phasors (3D over 3D with curl).}}
\label{fig:CurlvsNoCurlvs1D}
\end{center}
\end{figure}

\begin{figure}
\begin{center}
\includegraphics[width=0.45\textwidth] {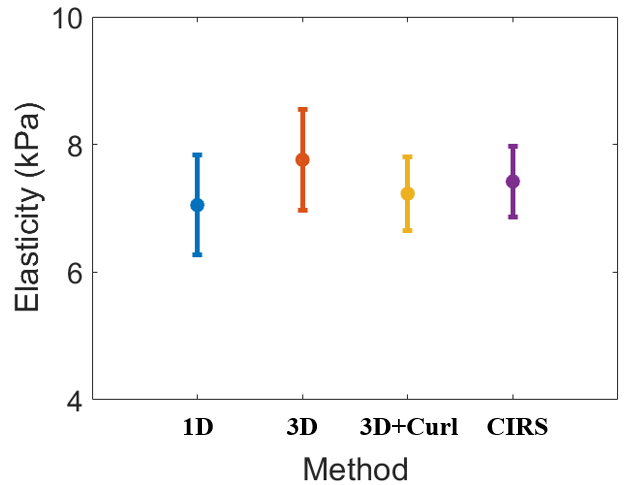}
\caption{\textcolor{black}{Comparison between different approaches in elasticity reconstruction for the 7.8 kPa phantom excited at 350 Hz. 
"1D" shows the 3D reconstruction using only the axial displacement phasor volume. "3D" shows the 3D reconstruction using all axial, lateral and elevational displacement volumes. "3D+curl" is a 3D reconstruction where the curl of the three displacement volumes (axial, lateral and elevational) is used in the reconstruction. "CIRS" shows the reported elasticity value by the manufacturer.}}
\label{fig:withCurlComparison}
\end{center}
\end{figure}

\textcolor{black}{Figure~\ref{fig:CurlvsNoCurlvs1D} shows 3D reconstruction results with different approaches for the 8 kPa phantom and an excitation frequency of 350 Hz. The CIRS reported elasticity value is $7.42 \pm 0.55$~kPa. The result of the 3D reconstruction using only the axial phasor volume is shown in (a) with the estimated elasticity value of $7.05 \pm 0.78$~kPa. (b) shows the results of the 3D reconstruction using axial, lateral and elevational displacement phasors (without using the curl) with an elasticity value of $7.76 \pm 0.79$~kPa. 
The result of 3D reconstruction with curl is shown in (c) with the estimated elasticity value of $7.23 \pm 0.58$~kPa. The STD value shows 25\% improvement compared to the results reported in (a) and (b), where the curl of the displacements has not been used. 
Furthermore, using curl of 3D displacement phasors provides closer elasticity value to the CIRS manufacturer reported value with less artifacts in the elasticity volume compared to 1D displacement estimation and also 3D displacements without using the curl. 
All the estimated elasticity and STD values are shown in Figure~\ref{fig:withCurlComparison}.}

\subsection{Heterogeneous Phantom}

\begin{figure*}
\begin{center}
\subfigure[inclusion 1 (PW)]{\includegraphics[width=0.35\textwidth] {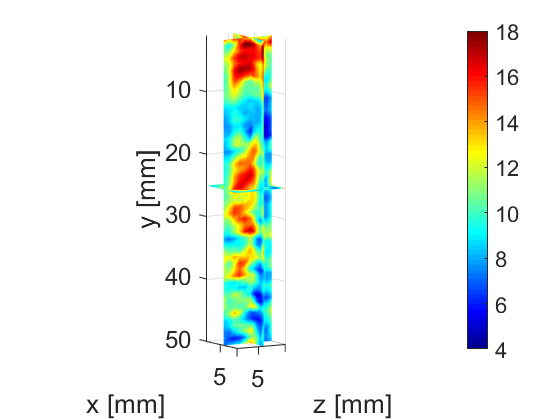}}
\subfigure[inclusion 1 (CDW)]{\includegraphics[width=0.35\textwidth] {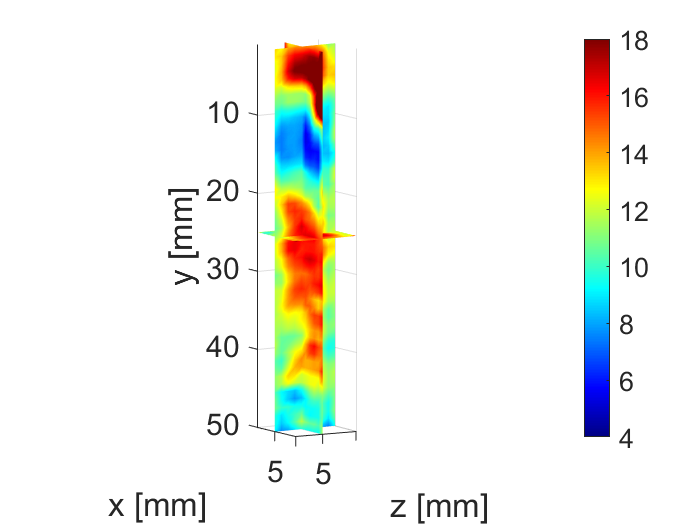}}

\subfigure[inclusion 2 (PW)]{\includegraphics[width=0.35\textwidth] {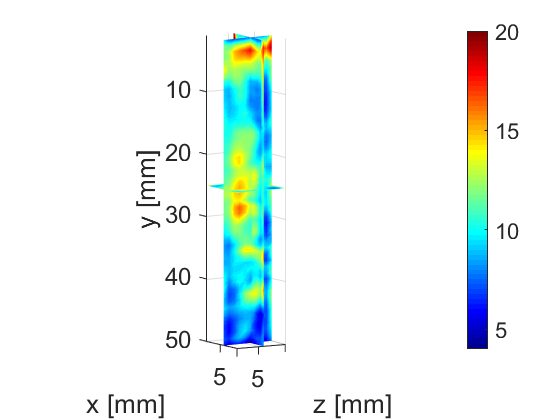}}
\subfigure[inclusion 2 (CDW)]{\includegraphics[width=0.35\textwidth] {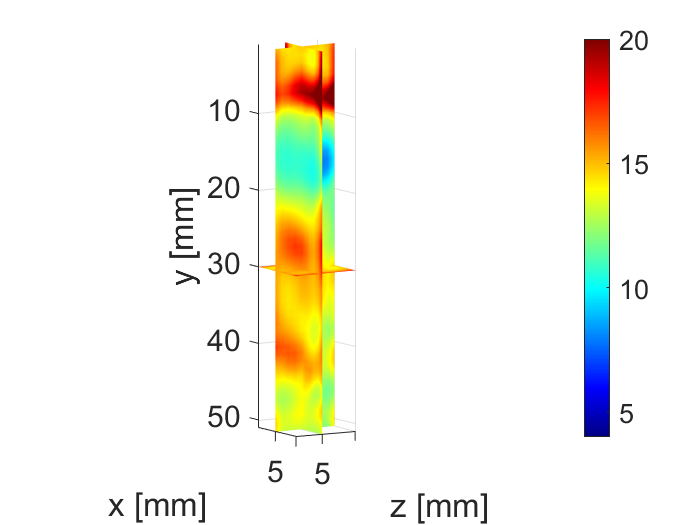}}

\subfigure[inclusion 3 (PW)]{\includegraphics[width=0.35\textwidth] {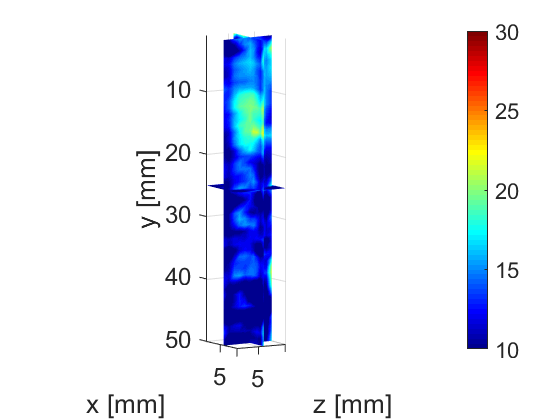}}
\subfigure[inclusion 3 (CDW)]{\includegraphics[width=0.35\textwidth] {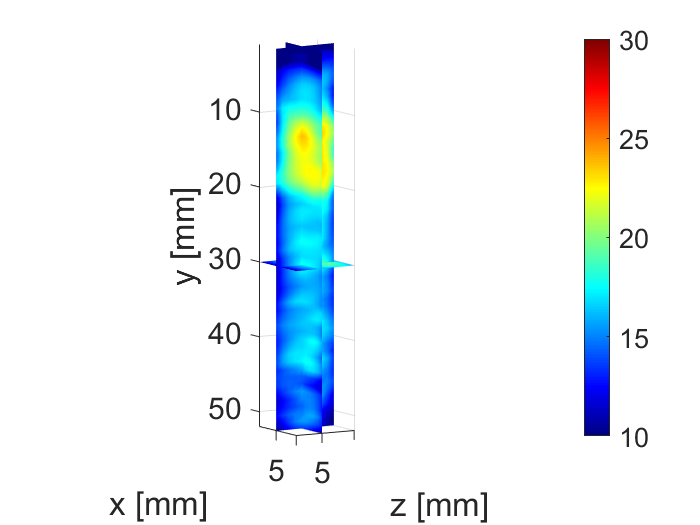}}

\subfigure[inclusion 4 (PW)]{\includegraphics[width=0.35\textwidth] {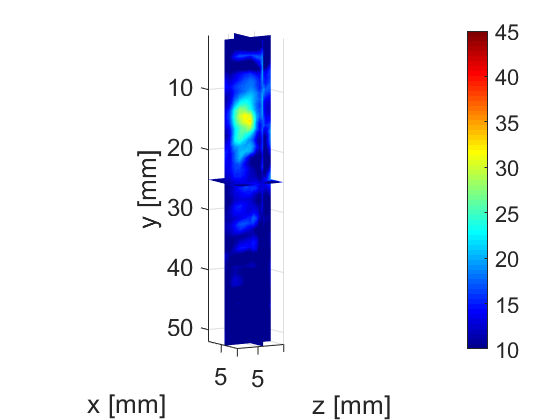}}
\subfigure[inclusion 4 (CDW)]{\includegraphics[width=0.35\textwidth] {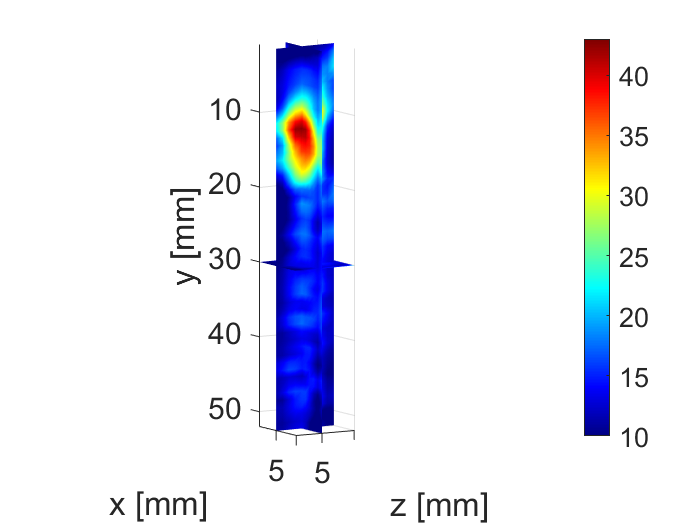}}
\caption{\textcolor{black}{Elasticity volumes for different inclusions inside of the heterogeneous phantom using a 400~Hz excitation frequency. \textcolor{black}{(a),(c),(e),(g) show the results of the plane wave (PW) imaging sequence, and (b),(d),(f),(h) demonstrate the results of the compounded diverging wave (CDW) imaging sequence.} The spherical inclusions are located at the depth of 10~mm to 20~mm with a diameter of 10~mm (same as the transducer footprint size). The value bar is in [kPa].}}
\label{fig:E_heterPhantom_volumes}
\end{center}
\end{figure*}

\begin{figure}
\begin{center}
\subfigure[Data volume]{\includegraphics[width=0.28\textwidth] {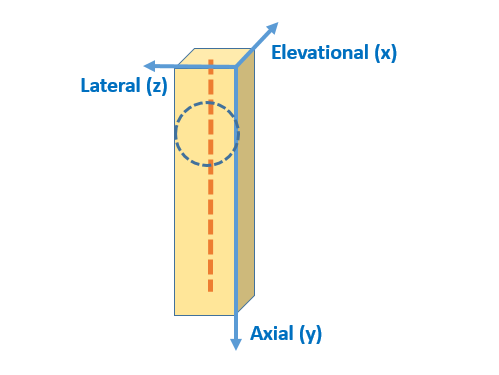}}
\subfigure[inclusion 1]{\includegraphics[width=0.39\textwidth] {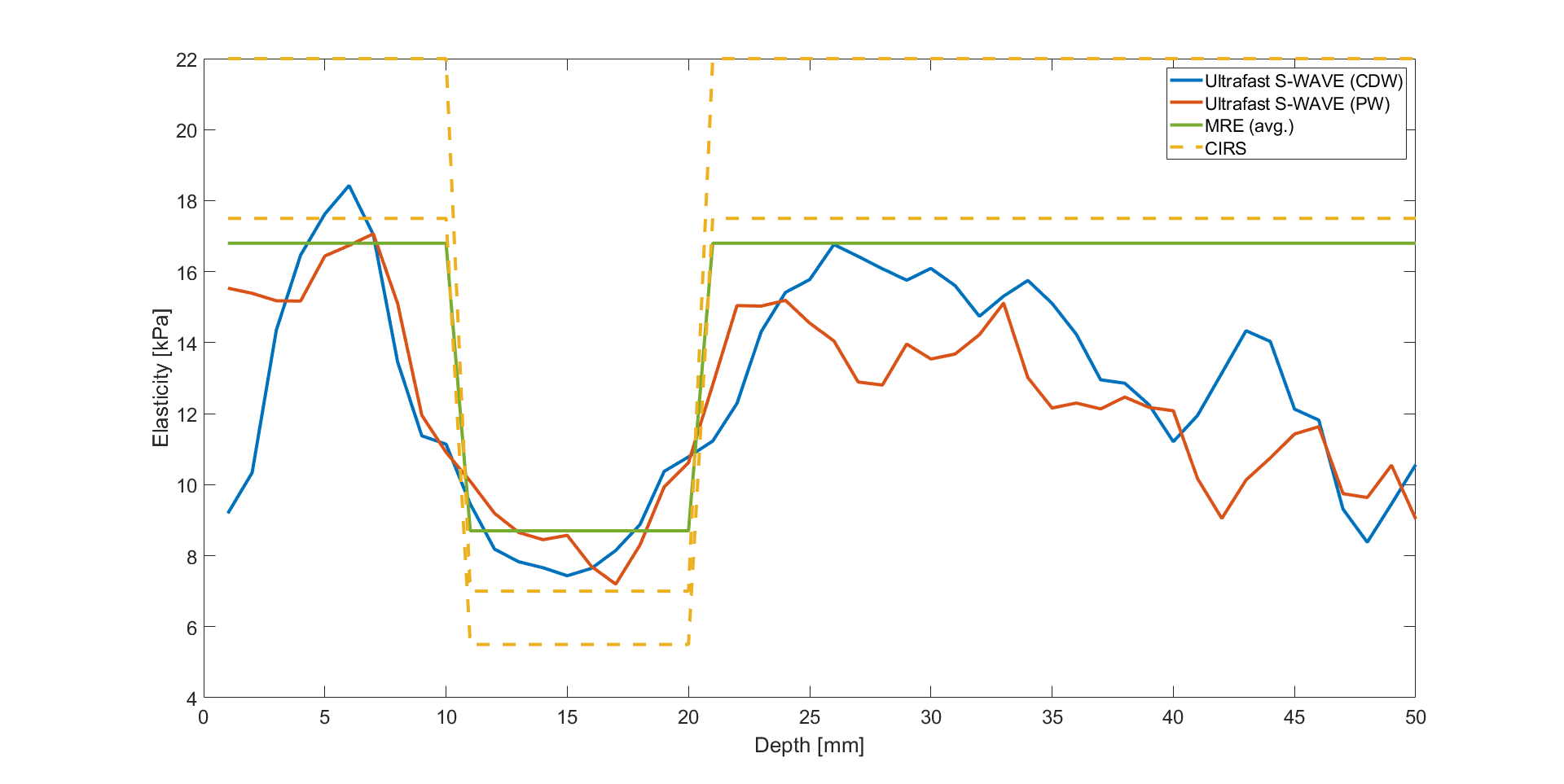}}
\subfigure[inclusion 2]{\includegraphics[width=0.39\textwidth] {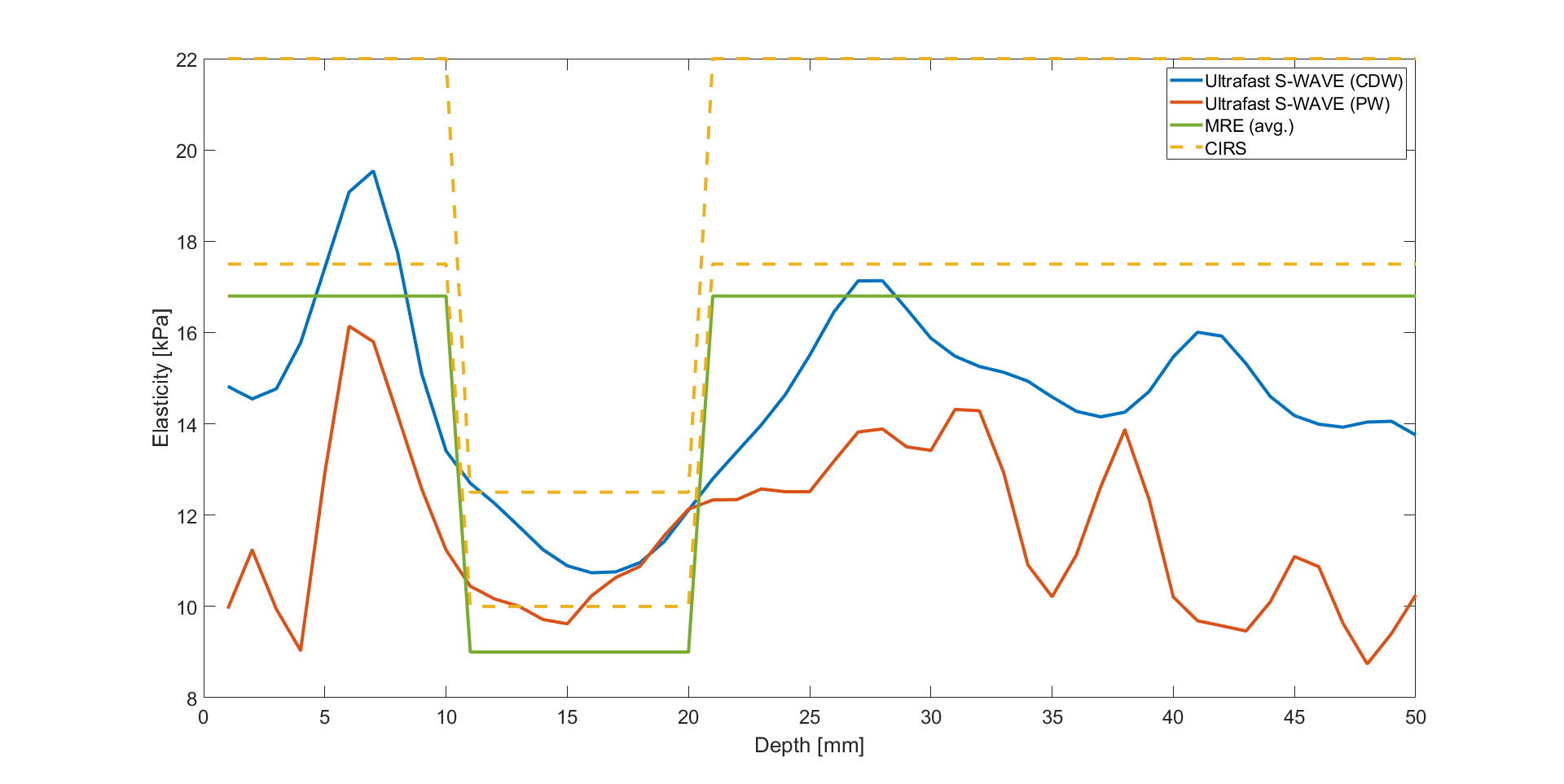}}
\subfigure[inclusion 3]{\includegraphics[width=0.39\textwidth] {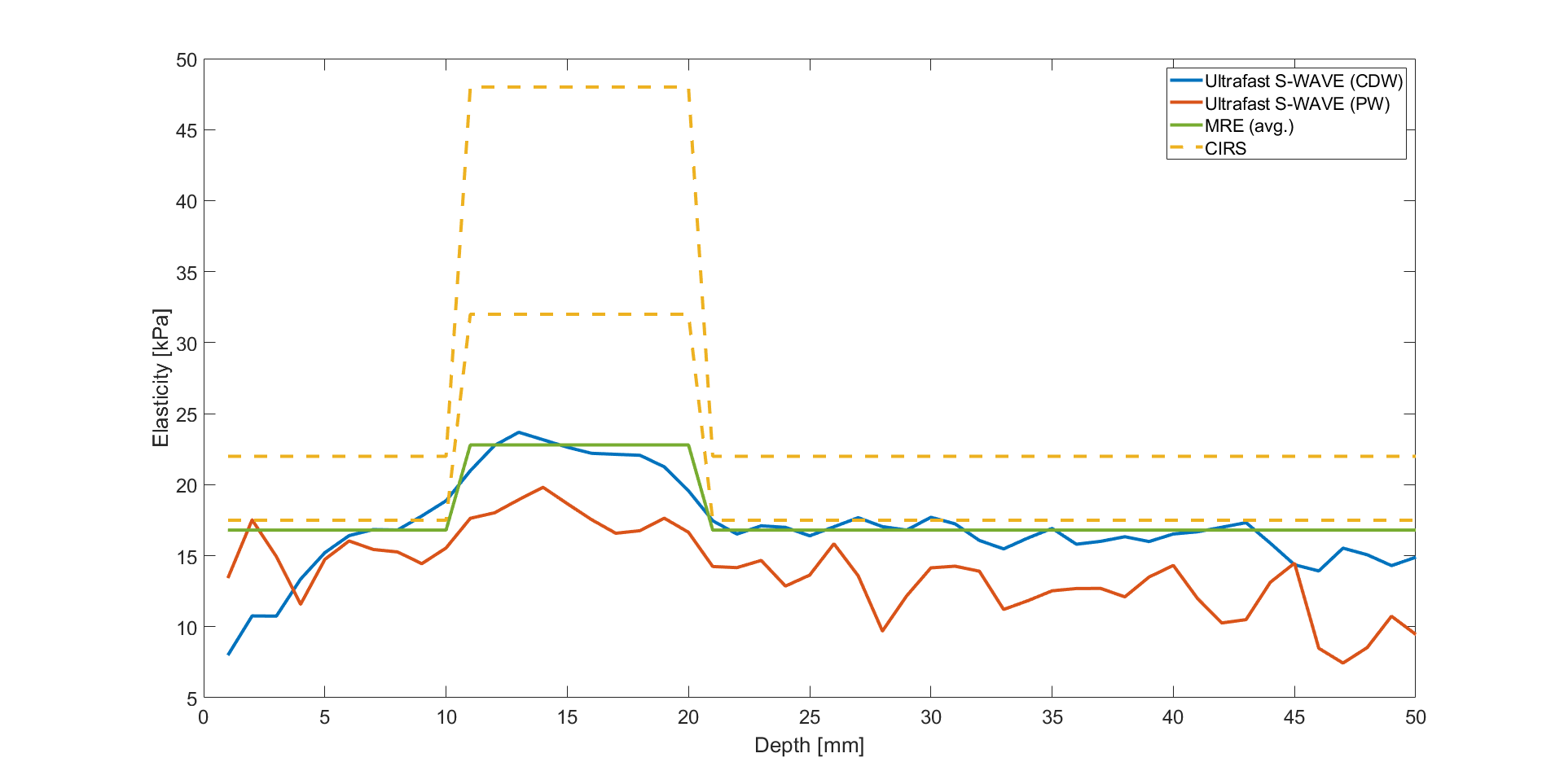}}
\subfigure[inclusion 4]{\includegraphics[width=0.39\textwidth] {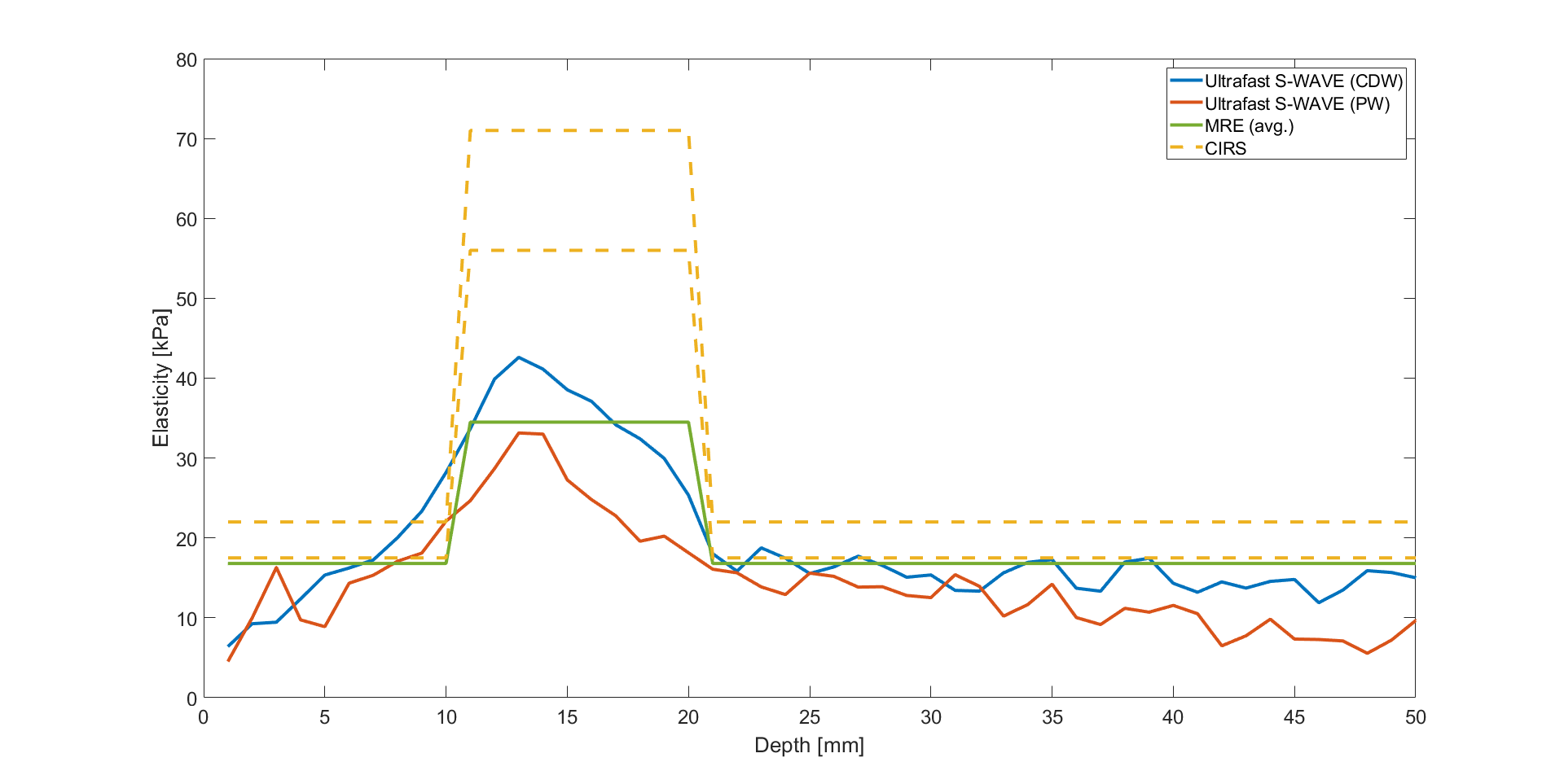}}
\caption{\textcolor{black}{Elasticity values of the midline of the volume versus the imaging depth \textcolor{black}{for both plane wave (PW) and compounded diverging wave (CDW) imaging sequences}. (a) shows the data volume, 3D coordinates of the transducer, and the location of the inclusion at the depth of 10~mm to 20~mm.(b)-(e) show the results for inclusions 1, 2, 3 and 4 at an excitation frequency of 400 Hz, respectively. The manufacturer elasticity range is shown with dashed lines. The average values by MRE are plotted in green. For each inclusion, the results of the compounded diverging waves sequence as well as an averaged curve over a 1~cm window in axial direction are presented.}}
\label{fig:E_heterPhantom_midline}
\end{center}
\end{figure}

\textcolor{black}{The matrix array transducer was placed on top of the phantom where each inclusion is located. For all the inclusions \textcolor{black}{both the PW and CDW imaging sequences} were used, as explained in~\ref{subsec:ImagingSequences}, to collect volumetric data. The elasticity volume for each excitation frequency was calculated and the elasticity values of a \textcolor{black}{$2\times2\times2$~mm$^3$ ROI at the center of the inclusion and inside} the background are averaged and shown along the STD values in Table~\ref{tab:E_heterPhantom}. 
For the Ultrafast S-WAVE method, the background elasticity in the last row of the table is an average value from all the volumes calculated in upper rows of the same column. \textcolor{black}{For the PW imaging sequence, the average error between the estimated values and the MRE values is 9\% with a maximum error of 14\%. For the CDW sequence, the average error between the estimated values and the MRE values is 6\% with a maximum error of 13\%. The results of the CDW sequence are closer to the MRE results.} 
The elasticity volumes for all inclusions at the excitation frequency of 400~Hz are shown in Figure~\ref{fig:E_heterPhantom_volumes} where all the inclusions are \textcolor{black}{located at the depth of 10-20~cm and} visible in the elasticity volumes. Figure~\ref{fig:E_heterPhantom_midline} (a) shows the imaging volume, the imaging coordinates, and the inclusion marked with a dashed circle and located at the depth of 10~mm to 20~mm. (b)-(e) show the elasticity values along} \textcolor{black}{the midline of the elasticity volume for different inclusions at a 400~Hz excitation frequency. For each inclusion, the results of the \textcolor{black}{PW and} CDW sequences are presented. The minimum and maximum elasticity values reported by the manufacturer are shown with dashed lines. The average elasticity value reported for the same phantom by MRE at the excitation frequencies of 200-250~Hz are also plotted~\cite{mohammed2022model}. 
The estimated values by the \textcolor{black}{CDW sequence are closer to the MRE values than those obtaiend with the PW sequence. Furthermore, both of the} proposed methods follow the MRE values closer compared to the values reported by the manufacturer. The same discrepancy has also been reported by previous studies using the same phantom model~\cite{baghani2011travelling, honarvar2013curl}. There are several potential explanations for this inconsistency, including variances in temperature, excitation frequency, or changes in the properties of the phantom material that may occur as it ages.} 

\begin{table*}[tb]
\centering
\caption{\textcolor{black}{The elasticity$\pm$STD values of the heterogeneous phantom \textcolor{black}{for an excitation frequency of 400~Hz}. \textcolor{black}{A $2$~mm$\times2$~mm$\times2$~mm window at the center of the inclusion and also inside the background is used for elasticity calculations}. The background elasticity is the average value of the elasticity volumes calculated in each column.}} 
\label{tab:E_heterPhantom}
\begin{tabular}{c c c c c c} 
\multicolumn{1}{c}{} &
\multicolumn{1}{c}{Manufact.} &
\multicolumn{1}{c}{MRE} &
\multicolumn{1}{c}{Ultrafast S-WAVE (PW)} &
\multicolumn{1}{c}{Ultrafast S-WAVE (CDW)}\\
\hline
Inc. 1 & 5.5-7 & 8.7$\pm$0.3 & 8.4$\pm$0.5 & 7.8$\pm$1.2 \\
Inc. 2 &  10-12.5 & 9.1$\pm$0.3 & 9.5$\pm$1.0 & 10.3$\pm$0.4 \\
Inc. 3 & 32-41 & 22.8$\pm$0.6 & 19.5$\pm$0.8 & 22.9$\pm$0.7 \\
Inc. 4 &  56-71 & 34.5$\pm$0.9 & 31.5$\pm$2.4 & 34.5$\pm$2.1 \\
Bkg. &  17.5-22 & 16.8$\pm$0.6 & 14.5$\pm$1.4 & 15.5$\pm$1.0 \\
\end{tabular}
\end{table*}

\subsection{Ex Vivo Study}

\begin{figure*}
\centerline{\includegraphics[width=1.9\columnwidth]{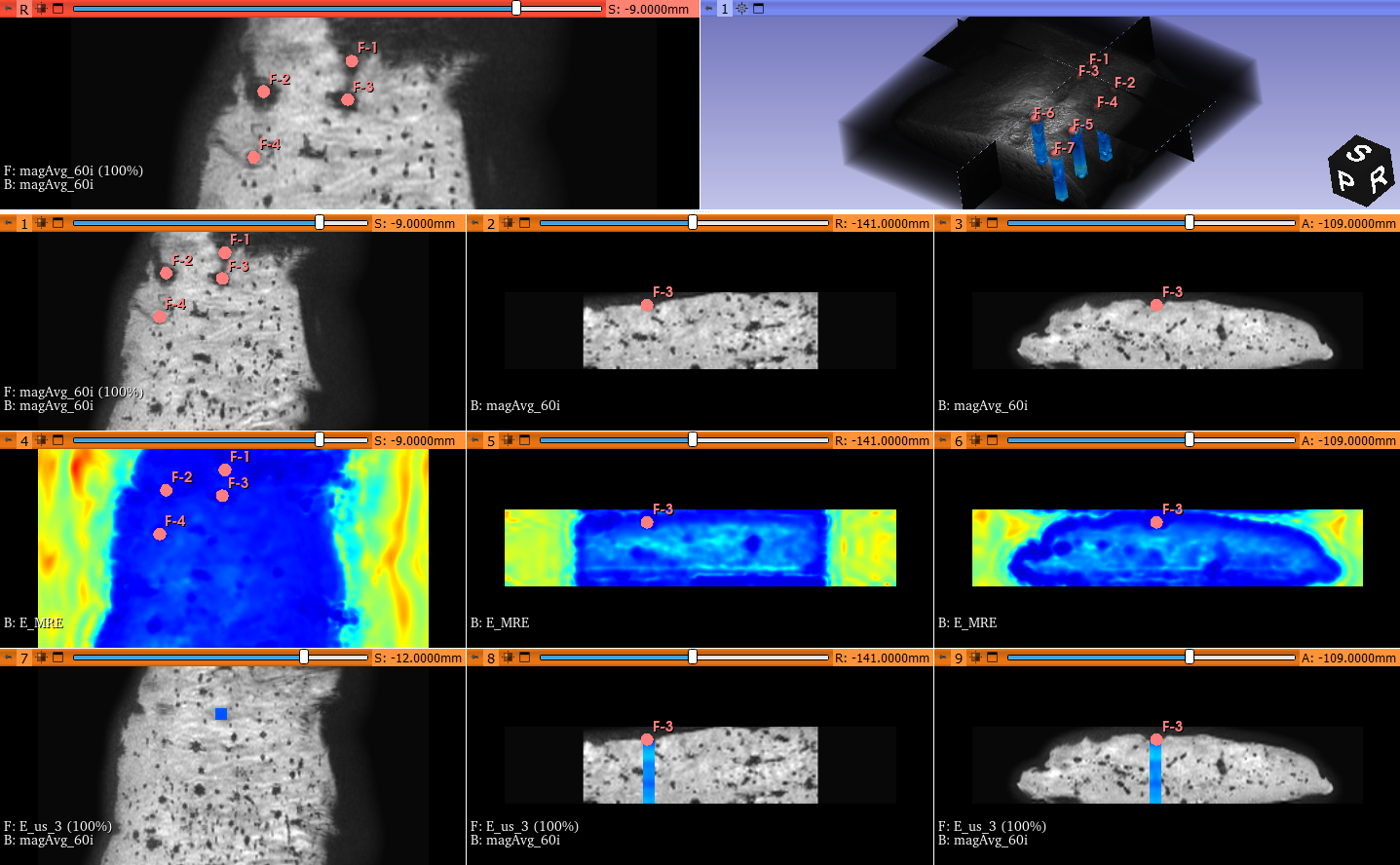}}
\caption{\textcolor{black}{Overlaying of MR T2 Weighted image and elasticity volumes. The first row (right image) shows the MR T2 weighted volume and Ultrafast S-WAVE elasticity volumes at seven measurement locations and in the left image, four (out of seven) measurement points are shown in the location of vitamin E capsules on a 2D slice of T2 weighted image. The second row depicts the same T2 weighted image volume in the axial, sagittal and coronal planes. The estimated MR elasticity is shown in the third row. The fourth row shows the overlay of the elasticity volume calculated by the Ultrafast S-WAVE method on the T2 weighted image volume for the measurement location 3.}}
\label{fig:slicerImg}
\end{figure*}

\begin{figure}[!t]
\centerline{\includegraphics[width=1\columnwidth]{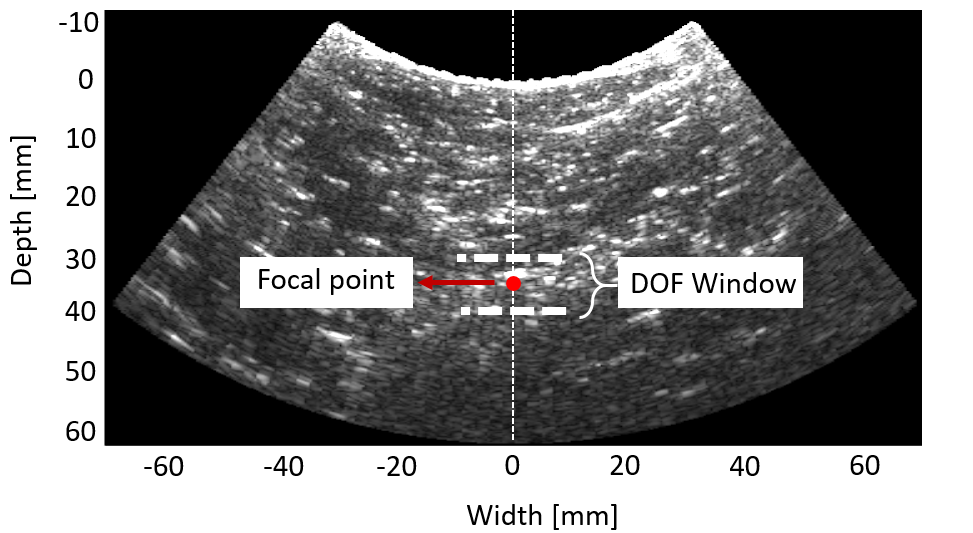}}
\caption{\textcolor{black}{ARFI ex vivo experiment. The B-mode image is captured using a C5-2v transducer before the start of the exam. The transducer is placed on each measurement location and the focal depth of the push beam is specified at the depth of 3.5~cm. The phase velocity estimation is performed by averaging over the axial DOF window, corresponding to an approximately 7.7 mm window at a frequency of 3.5 MHz, and defined by $8F^2\lambda$ around the focal depth.}}
\label{fig:exvivoARFIBmode}
\end{figure}

\textcolor{black}{The ex vivo bovine liver sample is imaged by three different methods: Ultrafast S-WAVE, MRE, and ARFI as explained in~\ref{subsubsec:method_exvivo}, \ref{subsec:MREValidation}, and \ref{subsec:ARFIValidation}. 
\textcolor{black}{For the Ultrafast S-WAVE method, we employ both the PW and CDW imaging sequences.}
Figure~\ref{fig:slicerImg} shows elasticity volumes overlayed onto T2 weighted images.} \textcolor{black}{The first row (right image) presents the T2 weighted image volume, seven measurement locations and Ultrafast S-WAVE elasticity volumes. Four (out of seven) measurement locations are shown in the first row (left image) where vitamin E capsules are also visible in the image at measurement locations. 
The second row shows the same T2 weighted image in axial, sagittal and coronal planes. The third row depicts the elasticity maps estimated by the MRE method. The last row presents the aligned elasticity volume of the Ultrafast S-WAVE as overlays on the T2 weighted images for the measurement location 3.} 
\textcolor{black}{Figure~\ref{fig:exvivoARFIBmode} shows the B-mode image right before the start of the ARFI exam. The transducer is placed on each marked locations on the sample and the acoustic radiation force is generated at the focal depth of the push beam which is located at the depth of 3.5~cm as shown in Figure~\ref{fig:exvivoARFIBmode}}.

\begin{figure}[!t]
\centerline{\includegraphics[width=1\columnwidth]{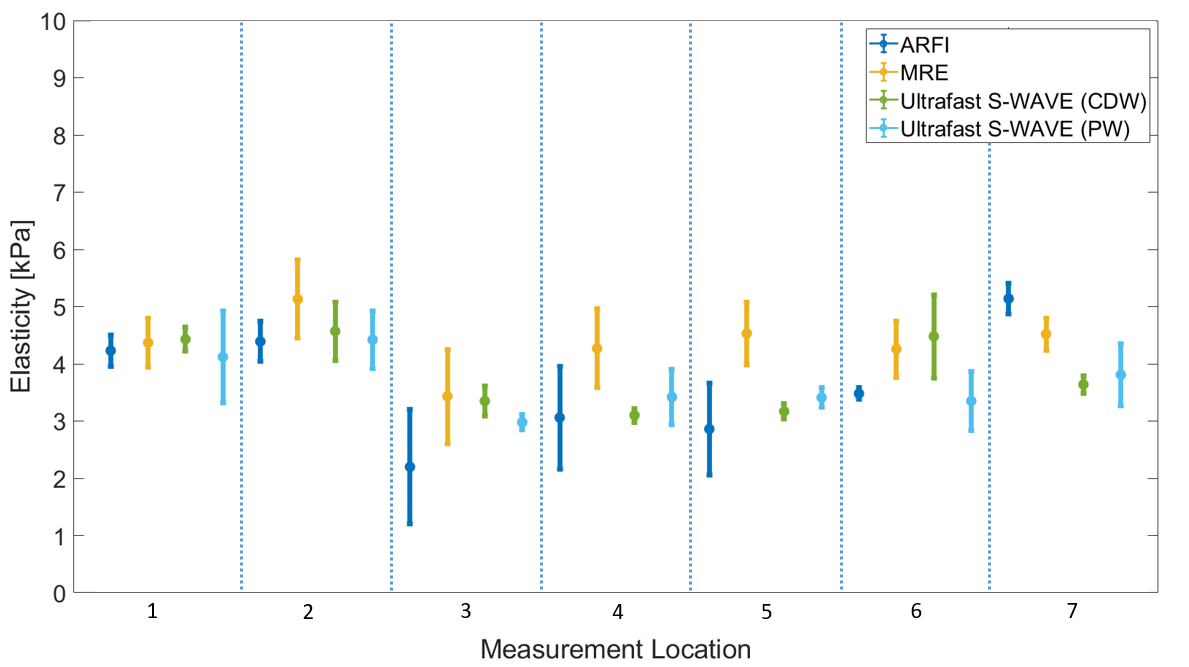}}
\caption{\textcolor{black}{Elasticity of the bovine liver sample by different elastography methods for seven different locations marked on the sample. The 5 mm registration error between acquisitions with different methods also contributes to the difference between measurements.}}
\label{fig:exvivoPlot}
\end{figure}

\textcolor{black}{The average elasticity as well as the STD values for all methods are shown in Figure~\ref{fig:exvivoPlot}} \textcolor{black}{for all seven measurement locations. In some measurement locations, the estimated values by different methods are slightly different due to proximity to a blood vessel or 5 mm registration error between acquisitions with different methods. However, all methods show approximately the same range for the elasticity values of the bovine liver tissue: \textcolor{black}{3.0-4.5 kPa by the Ultrafast S-WAVE (PW)}, 3.1-4.6 kPa by the Ultrafast S-WAVE (CDW), 3.4-5.1 kPa by MRE, and 2.9-4.4 kPa by the ARFI method. There is less than 9\% difference between the range of elasticity values provided by Ultrafast S-WAVE \textcolor{black}{(CDW sequence)} and the range of elasticity values provided by MRE and ARFI.} \textcolor{black}{The corresponding difference between the range of elasticity values achieved by the Ultrafast S-WAVE (PW sequence) and the range of elasticity values obtained by MRE and ARFI is less than 11\%.}

\section{Discussion}
\label{sec:discussion}

\textcolor{black}{This work presents a novel \textit{real-time} \textcolor{black}{volumetric data acquisition} based on the S-WAVE technique. As the tissue moves in all three directions after applying a force or vibration, measuring 3D displacement field over a volume improves the accuracy of the estimated elasticity values compared to the methods which measure only the axial displacement over a volume of data~\cite{hashemi20203d}.
To enable full potential of the proposed method where all three-directional phasors are utilized, it is recommended to have at least one full spatial wavelength of the shear waves in axial, lateral, and elevational directions.}
\textcolor{black}{For the matrix array transducer used in this work, the data length in the lateral and elevational directions is limited to the transducer size of 1~cm. To use Equation~\ref{eq:spatialWaveLength}, we should adjust the excitation frequency as a function of the expected phantom elasticity to have a spatial wavelength less than 1~cm. For example, for a phantom with elasticity less than 5 kPa, excitation frequencies $f_e > 120$~Hz give the spatial wavelength $\lambda < 1$~cm.}

\textcolor{black}{The curl of the 3D displacement field is used in elasticity reconstruction algorithm to cancel out the effect of the compressional waves\textcolor{black}{~\cite{baghani2009theoretical}} which was previously introduced in MRE \textcolor{black}{~\cite{sinkus2005viscoelastic,glaser2012review,manduca2018waveguide}}. It was also shown in a previous study~\cite{hashemi2022ultrafast} that using the 3D curl of displacements results in a more accurate elasticity value as well as a lower STD within homogeneous regions. However, the spatial derivatives inside the curl operation can be affected by the noisy displacement map. The GLUE3D motion estimation algorithm used in this work provides smooth displacements which are robust to noise~\cite{hashemi2017global} thereby alleviates the noise of the derivatives inside the curl operation. As an alternative to the curl operation, spatial high-pass filters uses the difference in propagation speed of compressional and shear waves to reduce the effects of the compressional waves. The drawback of this method is that the compression artifacts cannot be fully decoupled from the shear waves, and therefore, affect the shear wave speed estimation especially when the time-of-flight or phase gradient techniques are used~\cite{palmeri2008quantifying}.}

\textcolor{black}{In the \textit{ex vivo} experiment, the elasticity values are measured at seven different locations on the bovine liver sample showing a range of values. In previous studies on \textit{ex vivo} liver samples, a similar range of values has been reported for healthy liver tissue~\cite{barry2015shear,ormachea2016shear}.}

\textcolor{black}{In this work, the data processing is performed offline after collecting RF data. The beamforming has been implemented on GPUs and it takes 0.03~s for a volume of size $7\times1\times1$~cm$^3$. Phasor fitting, curl calculations and elasticity estimation functions have been implemented in C++ and can be used in MATLAB as mex functions. They are used in sequence to calculate the elasticity map from the displacements in 0.3~s. The displacement estimation has only been implemented in MATLAB. It can benefit from a GPU implementation in future work. For a volume of size $7\times1\times1$~cm$^3$, our current implementation of the displacement estimation takes 2 minutes to calculate the displacement maps between every two volumes of the mentioned size on a 3.6-GHz Intel Core i7 computer. Our future work includes integrating all the data processing steps to the Verasonics pipeline and using GPU implementations for the displacement estimation to work in real-time.}

\section{Conclusion}
\label{sec:conclusion}

\textcolor{black}{The proposed work aims to facilitate transmission of ultrasound diagnostic care being provided at hospitals to the point-of-care where the patients are located. This paper introduced the first \textit{real-time} 3D imaging of the S-WAVE method. The proposed technique provides a final number as the elasticity of the tissue which can be used by any ultrasound operator as an indication of the tissue condition. A matrix array transducer was connected to the Verasonics ultrasound machine to collect 100 RF volumes at the frame rates of 2000 volumes/s. The overall acquisition time is 0.05 s which has been substantially improved compared to the exam time in previous S-WAVE studies (12 s and 2 s). \textcolor{black}{Two imaging sequences are proposed: plane wave imaging, and compounded diverging wave imaging. The method validation on three liver fibrosis phantoms shows less than 8\% and 5\% differences between the estimated elasticity values by the proposed plane wave sequence and compounded diverging wave sequence, respectively, compared to the manufacturer reported ones.} \textcolor{black}{The proposed plane wave and compounded diverging wave sequences are applied to various inclusions within a heterogeneous phantom and the estimated elasticity values are compared with the provided values by MRE which show average errors of 9\% and 6\%, respectively.}
An \textit{ex vivo} study on a bovine liver sample showed \textcolor{black}{less than 9\% difference in the range of values provided by the} proposed method, MRE, and ARFI. Future work will examine the proposed method on patients with liver fibrosis disease.}

\section*{Acknowledgment}
\textcolor{black}{The ultrasound data was collected at DarkVision Technologies Inc. We thank Dr. Vincent Perrot for valuable discussions and the anonymous reviewers for their constructive feedback.}

\textcolor{black}{
\bibliographystyle{IEEEtran}
\bibliography{ref}}

\begin{thebibliography}{10}
\providecommand{\url}[1]{#1}
\csname url@samestyle\endcsname
\providecommand{\newblock}{\relax}
\providecommand{\bibinfo}[2]{#2}
\providecommand{\BIBentrySTDinterwordspacing}{\spaceskip=0pt\relax}
\providecommand{\BIBentryALTinterwordstretchfactor}{4}
\providecommand{\BIBentryALTinterwordspacing}{\spaceskip=\fontdimen2\font plus
\BIBentryALTinterwordstretchfactor\fontdimen3\font minus
  \fontdimen4\font\relax}
\providecommand{\BIBforeignlanguage}[2]{{%
\expandafter\ifx\csname l@#1\endcsname\relax
\typeout{** WARNING: IEEEtran.bst: No hyphenation pattern has been}%
\typeout{** loaded for the language `#1'. Using the pattern for}%
\typeout{** the default language instead.}%
\else
\language=\csname l@#1\endcsname
\fi
#2}}
\providecommand{\BIBdecl}{\relax}
\BIBdecl

\bibitem{liu2009real}
D.~Liu and E.~S. Ebbini, ``Real-time 2-d temperature imaging using
  ultrasound,'' \emph{IEEE Transactions on Biomedical Engineering}, vol.~57,
  no.~1, pp. 12--16, 2009.

\bibitem{lewis2015thermometry}
M.~A. Lewis, R.~M. Staruch, and R.~Chopra, ``Thermometry and ablation
  monitoring with ultrasound,'' \emph{International Journal of Hyperthermia},
  vol.~31, no.~2, pp. 163--181, 2015.

\bibitem{brock2012impact}
M.~Brock, C.~von Bodman, R.~J. Palisaar, B.~L{\"o}ppenberg, F.~Sommerer,
  T.~Deix, J.~Noldus, and T.~Eggert, ``The impact of real-time elastography
  guiding a systematic prostate biopsy to improve cancer detection rate: a
  prospective study of 353 patients,'' \emph{The Journal of Urology}, vol. 187,
  no.~6, pp. 2039--2043, 2012.

\bibitem{moe2020transrectal}
A.~Moe and D.~Hayne, ``Transrectal ultrasound biopsy of the prostate: does it
  still have a role in prostate cancer diagnosis?'' \emph{Translational
  Andrology and Urology}, vol.~9, no.~6, p. 3018, 2020.

\bibitem{miyagawa2010real}
T.~Miyagawa, S.~Ishikawa, T.~Kimura, T.~Suetomi, M.~Tsutsumi, T.~Irie,
  M.~Kondoh, and T.~Mitake, ``Real-time virtual sonography for navigation
  during targeted prostate biopsy using magnetic resonance imaging data,''
  \emph{International Journal of Urology}, vol.~17, no.~10, pp. 855--860, 2010.

\bibitem{valerio2015detection}
M.~Valerio, I.~Donaldson, M.~Emberton, B.~Ehdaie, B.~A. Hadaschik, L.~S. Marks,
  P.~Mozer, A.~R. Rastinehad, and H.~U. Ahmed, ``Detection of clinically
  significant prostate cancer using magnetic resonance imaging--ultrasound
  fusion targeted biopsy: a systematic review,'' \emph{European Urology},
  vol.~68, no.~1, pp. 8--19, 2015.

\bibitem{pinto2011magnetic}
P.~A. Pinto, P.~H. Chung, A.~R. Rastinehad, A.~A. Baccala, J.~Kruecker, C.~J.
  Benjamin, S.~Xu, P.~Yan, S.~Kadoury, C.~Chua \emph{et~al.}, ``Magnetic
  resonance imaging/ultrasound fusion guided prostate biopsy improves cancer
  detection following transrectal ultrasound biopsy and correlates with
  multiparametric magnetic resonance imaging,'' \emph{The Journal of Urology},
  vol. 186, no.~4, pp. 1281--1285, 2011.

\bibitem{hyun2020nondestructive}
D.~Hyun, L.~Abou-Elkacem, R.~Bam, L.~L. Brickson, C.~D. Herickhoff, and J.~J.
  Dahl, ``Nondestructive detection of targeted microbubbles using dual-mode
  data and deep learning for real-time ultrasound molecular imaging,''
  \emph{IEEE Transactions on Medical Imaging}, vol.~39, no.~10, pp. 3079--3088,
  2020.

\bibitem{perera2020real}
R.~H. Perera, X.~Wang, Y.~Wang, G.~Ramamurthy, P.~Peiris, E.~Abenojar, J.~P.
  Basilion, A.~A. Exner \emph{et~al.}, ``Real time ultrasound molecular imaging
  of prostate cancer with psma-targeted nanobubbles,'' \emph{Nanomedicine:
  Nanotechnology, Biology and Medicine}, vol.~28, p. 102213, 2020.

\bibitem{dodd2000minimally}
G.~D. Dodd, M.~C. Soulen, R.~A. Kane, T.~Livraghi, W.~R. Lees, Y.~Yamashita,
  A.~R. Gillams, O.~I. Karahan, and H.~Rhim, ``Minimally invasive treatment of
  malignant hepatic tumors: at the threshold of a major breakthrough,''
  \emph{Radiographics}, vol.~20, no.~1, pp. 9--27, 2000.

\bibitem{wi2015real}
H.~Wi, A.~L. McEwan, V.~Lam, H.~J. Kim, E.~J. Woo, and T.~I. Oh, ``Real-time
  conductivity imaging of temperature and tissue property changes during
  radiofrequency ablation: an ex vivo model using weighted frequency
  difference,'' \emph{Bioelectromagnetics}, vol.~36, no.~4, pp. 277--286, 2015.

\bibitem{bharat2005monitoring}
S.~Bharat, U.~Techavipoo, M.~Z. Kiss, W.~Liu, and T.~Varghese, ``Monitoring
  stiffness changes in lesions after radiofrequency ablation at different
  temperatures and durations of ablation,'' \emph{Ultrasound in Medicine \&
  Biology}, vol.~31, no.~3, pp. 415--422, 2005.

\bibitem{van2010intra}
M.~G. Van~Vledder, E.~M. Boctor, L.~R. Assumpcao, H.~Rivaz, P.~Foroughi, G.~D.
  Hager, U.~M. Hamper, T.~M. Pawlik, and M.~A. Choti, ``Intra-operative
  ultrasound elasticity imaging for monitoring of hepatic tumour thermal
  ablation,'' \emph{Hpb}, vol.~12, no.~10, pp. 717--723, 2010.

\bibitem{hashemi2017global}
H.~S. Hashemi and H.~Rivaz, ``Global time-delay estimation in ultrasound
  elastography,'' \emph{IEEE Transactions on Ultrasonics, Ferroelectrics, and
  Frequency Control}, vol.~64, no.~10, pp. 1625--1636, 2017.

\bibitem{van2014comparison}
A.~van Hove, P.-H. Savoie, C.~Maurin, S.~Brunelle, G.~Gravis, N.~Salem, and
  J.~Walz, ``Comparison of image-guided targeted biopsies versus systematic
  randomized biopsies in the detection of prostate cancer: a systematic
  literature review of well-designed studies,'' \emph{World Journal of
  Urology}, vol.~32, no.~4, pp. 847--858, 2014.

\bibitem{postema2015multiparametric}
A.~Postema, M.~Mischi, J.~De~La~Rosette, and H.~Wijkstra, ``Multiparametric
  ultrasound in the detection of prostate cancer: a systematic review,''
  \emph{World Journal of Urology}, vol.~33, no.~11, pp. 1651--1659, 2015.

\bibitem{abeysekera2015vibro}
J.~Abeysekera, R.~Rohling, and S.~Salcudean, ``Vibro-elastography: Absolute
  elasticity from motorized 3d ultrasound measurements of harmonic motion
  vectors,'' in \emph{2015 IEEE International Ultrasonics Symposium
  (IUS)}.\hskip 1em plus 0.5em minus 0.4em\relax IEEE, 2015, pp. 1--4.

\bibitem{abeysekera2017swave}
J.~M. Abeysekera, M.~Ma, M.~Pesteie, J.~Terry, D.~Pugash, J.~A. Hutcheon,
  C.~Mayer, L.~Lampe, S.~Salcudean, and R.~Rohling, ``Swave imaging of
  placental elasticity and viscosity: proof of concept,'' \emph{Ultrasound in
  Medicine \& Biology}, vol.~43, no.~6, pp. 1112--1124, 2017.

\bibitem{mohammed2022model}
S.~Mohammed, M.~Honarvar, Q.~Zeng, H.~Hashemi, R.~Rohling, P.~Kozlowski, and
  S.~Salcudean, ``Model-based quantitative elasticity reconstruction using
  admm,'' \emph{IEEE Transactions on Medical Imaging}, 2022.

\bibitem{zeng2020three}
Q.~Zeng, M.~Honarvar, C.~Schneider, S.~K. Mohammad, J.~Lobo, E.~H. Pang, K.~T.
  Lau, C.~Hu, J.~Jago, S.~R. Erb \emph{et~al.}, ``Three-dimensional
  multi-frequency shear wave absolute vibro-elastography (3d s-wave) with a
  matrix array transducer: implementation and preliminary in vivo study of the
  liver,'' \emph{IEEE Transactions on Medical Imaging}, vol.~40, no.~2, pp.
  648--660, 2020.

\bibitem{zeng2022multifrequency}
Q.~Zeng, S.~Mohammed, T.~A. Aleef, E.~H. Pang, C.~Hu, J.~Jago, R.~Rohling, and
  S.~E. Salcudean, ``Multifrequency liver shear wave absolute
  vibro-elastography with an xmatrix array-2d vs. 3d comparison study,'' in
  \emph{2022 IEEE International Ultrasonics Symposium (IUS)}.\hskip 1em plus
  0.5em minus 0.4em\relax IEEE, 2022, pp. 1--4.

\bibitem{riccabona2003potential}
M.~Riccabona, G.~Fritz, and E.~Ring, ``Potential applications of
  three-dimensional ultrasound in the pediatric urinary tract: pictorial
  demonstration based on preliminary results,'' \emph{European radiology},
  vol.~13, no.~12, pp. 2680--2687, 2003.

\bibitem{hashemi2022ultrafast}
H.~S. Hashemi, S.~E. Salcudean, and R.~N. Rohling, ``Ultrafast ultrasound
  imaging for 3d shear wave absolute vibro-elastography,'' \emph{arXiv preprint
  arXiv:2203.13949}, 2022.

\bibitem{loomba2013global}
R.~Loomba and A.~J. Sanyal, ``The global nafld epidemic,'' \emph{Nature reviews
  Gastroenterology \& hepatology}, vol.~10, no.~11, pp. 686--690, 2013.

\bibitem{bellentani2010epidemiology}
S.~Bellentani, F.~Scaglioni, M.~Marino, and G.~Bedogni, ``Epidemiology of
  non-alcoholic fatty liver disease,'' \emph{Digestive Diseases}, vol.~28,
  no.~1, pp. 155--161, 2010.

\bibitem{foucher2006diagnosis}
J.~Foucher, E.~Chanteloup, J.~Vergniol, L.~Castera, B.~Le~Bail, X.~Adhoute,
  J.~Bertet, P.~Couzigou, and V.~de~Ledinghen, ``Diagnosis of cirrhosis by
  transient elastography (fibroscan): a prospective study,'' \emph{Gut},
  vol.~55, no.~3, pp. 403--408, 2006.

\bibitem{shao2021breast}
Y.~Shao, H.~Hashemi, P.~Gordon, L.~Warren, Z.~J. Wang, R.~Rohling, and
  T.~Salcudean, ``Breast cancer detection using multimodal time series features
  from ultrasound shear wave absolute vibro-elastography,'' \emph{IEEE Journal
  of Biomedical and Health Informatics}, 2021.

\bibitem{nightingale2015derivation}
K.~R. Nightingale, N.~C. Rouze, S.~J. Rosenzweig, M.~H. Wang, M.~F. Abdelmalek,
  C.~D. Guy, and M.~L. Palmeri, ``Derivation and analysis of viscoelastic
  properties in human liver: impact of frequency on fibrosis and steatosis
  staging,'' \emph{IEEE Transactions on Ultrasonics, Ferroelectrics, and
  Frequency Control}, vol.~62, no.~1, pp. 165--175, 2015.

\bibitem{aichele2021fluids}
J.~Aichele and S.~Catheline, ``Fluids alter elasticity measurements: Porous
  wave propagation accounts for shear wave dispersion in elastography,''
  \emph{Frontiers in Physics}, p. 536, 2021.

\bibitem{deng2016ultrasonic}
Y.~Deng, N.~C. Rouze, M.~L. Palmeri, and K.~R. Nightingale, ``Ultrasonic shear
  wave elasticity imaging sequencing and data processing using a verasonics
  research scanner,'' \emph{IEEE Transactions on Ultrasonics, Ferroelectrics,
  and Frequency Control}, vol.~64, no.~1, pp. 164--176, 2016.

\bibitem{provost20143d}
J.~Provost, C.~Papadacci, J.~E. Arango, M.~Imbault, M.~Fink, J.-L. Gennisson,
  M.~Tanter, and M.~Pernot, ``3d ultrafast ultrasound imaging in vivo,''
  \emph{Physics in Medicine \& Biology}, vol.~59, no.~19, p.~L1, 2014.

\bibitem{papadacci2014high}
C.~Papadacci, M.~Pernot, M.~Couade, M.~Fink, and M.~Tanter, ``High-contrast
  ultrafast imaging of the heart,'' \emph{IEEE transactions on ultrasonics,
  ferroelectrics, and frequency control}, vol.~61, no.~2, pp. 288--301, 2014.

\bibitem{wilcox2018quantification}
P.~D. Wilcox and J.~Zhang, ``Quantification of the effect of array element
  pitch on imaging performance,'' \emph{IEEE transactions on ultrasonics,
  ferroelectrics, and frequency control}, vol.~65, no.~4, pp. 600--616, 2018.

\bibitem{perrot2021so}
V.~Perrot, M.~Polichetti, F.~Varray, and D.~Garcia, ``So you think you can das?
  a viewpoint on delay-and-sum beamforming,'' \emph{Ultrasonics}, vol. 111, p.
  106309, 2021.

\bibitem{grondin2017cardiac}
J.~Grondin, V.~Sayseng, and E.~E. Konofagou, ``Cardiac strain imaging with
  coherent compounding of diverging waves,'' \emph{IEEE Transactions on
  Ultrasonics, Ferroelectrics, and Frequency Control}, vol.~64, no.~8, pp.
  1212--1222, 2017.

\bibitem{hashemi20203d}
H.~S. Hashemi, M.~Honarvar, T.~Salcudean, and R.~Rohling, ``3d global
  time-delay estimation for shear-wave absolute vibro-elastography of the
  placenta,'' in \emph{2020 42nd Annual International Conference of the IEEE
  Engineering in Medicine \& Biology Society (EMBC)}.\hskip 1em plus 0.5em
  minus 0.4em\relax IEEE, 2020, pp. 2079--2083.

\bibitem{azar2010sub}
R.~Z. Azar, O.~Goksel, and S.~E. Salcudean, ``Sub-sample displacement
  estimation from digitized ultrasound rf signals using multi-dimensional
  polynomial fitting of the cross-correlation function,'' \emph{IEEE
  Transactions on Ultrasonics, Ferroelectrics, and Frequency Control}, vol.~57,
  no.~11, pp. 2403--2420, 2010.

\bibitem{abeysekera2016three}
J.~M. Abeysekera, ``Three dimensional ultrasound elasticity imaging,'' Ph.D.
  dissertation, University of British Columbia, 2016.

\bibitem{knutsson1994local}
H.~Knutsson, C.-F. Westin, and G.~H. Granlund, ``Local multiscale frequency and
  bandwidth estimation.'' in \emph{ICIP (1)}, 1994, pp. 36--40.

\bibitem{manduca1996image}
A.~Manduca, R.~Muthupillai, P.~J. Rossman, J.~F. Greenleaf, and R.~L. Ehman,
  ``Image processing for magnetic-resonance elastography,'' in \emph{Medical
  Imaging 1996: Image Processing}, vol. 2710.\hskip 1em plus 0.5em minus
  0.4em\relax SPIE, 1996, pp. 616--623.

\bibitem{deffieux2011effects}
T.~Deffieux, J.-L. Gennisson, J.~Bercoff, and M.~Tanter, ``On the effects of
  reflected waves in transient shear wave elastography,'' \emph{IEEE
  Transactions on Ultrasonics, Ferroelectrics, and Frequency Control}, vol.~58,
  no.~10, pp. 2032--2035, 2011.

\bibitem{ormachea2016shear}
J.~Ormachea, R.~J. Lavarello, S.~A. McAleavey, K.~J. Parker, and B.~Castaneda,
  ``Shear wave speed measurements using crawling wave sonoelastography and
  single tracking location shear wave elasticity imaging for tissue
  characterization,'' \emph{IEEE Transactions on Ultrasonics, Ferroelectrics,
  and Frequency Control}, vol.~63, no.~9, pp. 1351--1360, 2016.

\bibitem{kikinis20143d}
R.~Kikinis, S.~D. Pieper, and K.~G. Vosburgh, ``3d slicer: a platform for
  subject-specific image analysis, visualization, and clinical support,'' in
  \emph{Intraoperative imaging and image-guided therapy}.\hskip 1em plus 0.5em
  minus 0.4em\relax Springer, 2014, pp. 277--289.

\bibitem{manduca2018waveguide}
A.~Manduca, T.~Rossman, D.~Lake, K.~Glaser, A.~Arani, S.~Arunachalam,
  P.~Rossman, J.~Trzasko, R.~Ehman, D.~Dragomir-Daescu \emph{et~al.},
  ``Waveguide effects and implications for cardiac magnetic resonance
  elastography: A finite element study,'' \emph{NMR in Biomedicine}, vol.~31,
  no.~10, p. e3996, 2018.

\bibitem{baghani2011travelling}
A.~Baghani, S.~Salcudean, M.~Honarvar, R.~S. Sahebjavaher, R.~Rohling, and
  R.~Sinkus, ``Travelling wave expansion: a model fitting approach to the
  inverse problem of elasticity reconstruction,'' \emph{IEEE Transactions on
  Medical Imaging}, vol.~30, no.~8, pp. 1555--1565, 2011.

\bibitem{honarvar2013curl}
M.~Honarvar, R.~Sahebjavaher, R.~Sinkus, R.~Rohling, and S.~E. Salcudean,
  ``Curl-based finite element reconstruction of the shear modulus without
  assuming local homogeneity: time harmonic case,'' \emph{IEEE transactions on
  medical imaging}, vol.~32, no.~12, pp. 2189--2199, 2013.

\bibitem{baghani2009theoretical}
A.~Baghani, S.~Salcudean, and R.~Rohling, ``Theoretical limitations of the
  elastic wave equation inversion for tissue elastography,'' \emph{The Journal
  of the Acoustical Society of America}, vol. 126, no.~3, pp. 1541--1551, 2009.

\bibitem{sinkus2005viscoelastic}
R.~Sinkus, M.~Tanter, T.~Xydeas, S.~Catheline, J.~Bercoff, and M.~Fink,
  ``Viscoelastic shear properties of in vivo breast lesions measured by mr
  elastography,'' \emph{Magnetic Resonance Imaging}, vol.~23, no.~2, pp.
  159--165, 2005.

\bibitem{glaser2012review}
K.~J. Glaser, A.~Manduca, and R.~L. Ehman, ``Review of mr elastography
  applications and recent developments,'' \emph{Journal of Magnetic Resonance
  Imaging}, vol.~36, no.~4, pp. 757--774, 2012.

\bibitem{palmeri2008quantifying}
M.~L. Palmeri, M.~H. Wang, J.~J. Dahl, K.~D. Frinkley, and K.~R. Nightingale,
  ``Quantifying hepatic shear modulus in vivo using acoustic radiation force,''
  \emph{Ultrasound in Medicine \& Biology}, vol.~34, no.~4, pp. 546--558, 2008.

\bibitem{barry2015shear}
C.~T. Barry, C.~Hazard, Z.~Hah, G.~Cheng, A.~Partin, R.~A. Mooney, K.-H.
  Chuang, W.~Cao, D.~J. Rubens, and K.~J. Parker, ``Shear wave dispersion in
  lean versus steatotic rat livers,'' \emph{Journal of Ultrasound in Medicine},
  vol.~34, no.~6, pp. 1123--1129, 2015.

\end{thebibliography}

\end{document}